\documentclass[a4paper,11pt]{article}
\pdfoutput=1 
\usepackage{jheppub} 
\usepackage[T1]{fontenc} 
\RequirePackage[displaymath]{lineno} 
\usepackage[percent]{overpic}
\usepackage{epsfig}
\usepackage{graphicx}
\usepackage{dcolumn}
\usepackage{bm}
\usepackage{ltablex,booktabs}
\usepackage{subfigure}
\usepackage{float}
\usepackage{color}
\usepackage{amsmath}
\usepackage{mathcomp}
\usepackage{mathrsfs}
\usepackage{multirow}
\usepackage{rotating}
\usepackage{amssymb}
\usepackage{gensymb}
\usepackage{tabularx}
\usepackage{epstopdf}
\usepackage{CJKutf8} 
\usepackage{comment}

\usepackage{siunitx} 
\title{Amplitude Analysis and Branching Fraction Measurement of $D^+ \to \pi^+\pi^0\pi^0$}
\collaborationImg{\includegraphics[width=0.15\textwidth, angle=90]{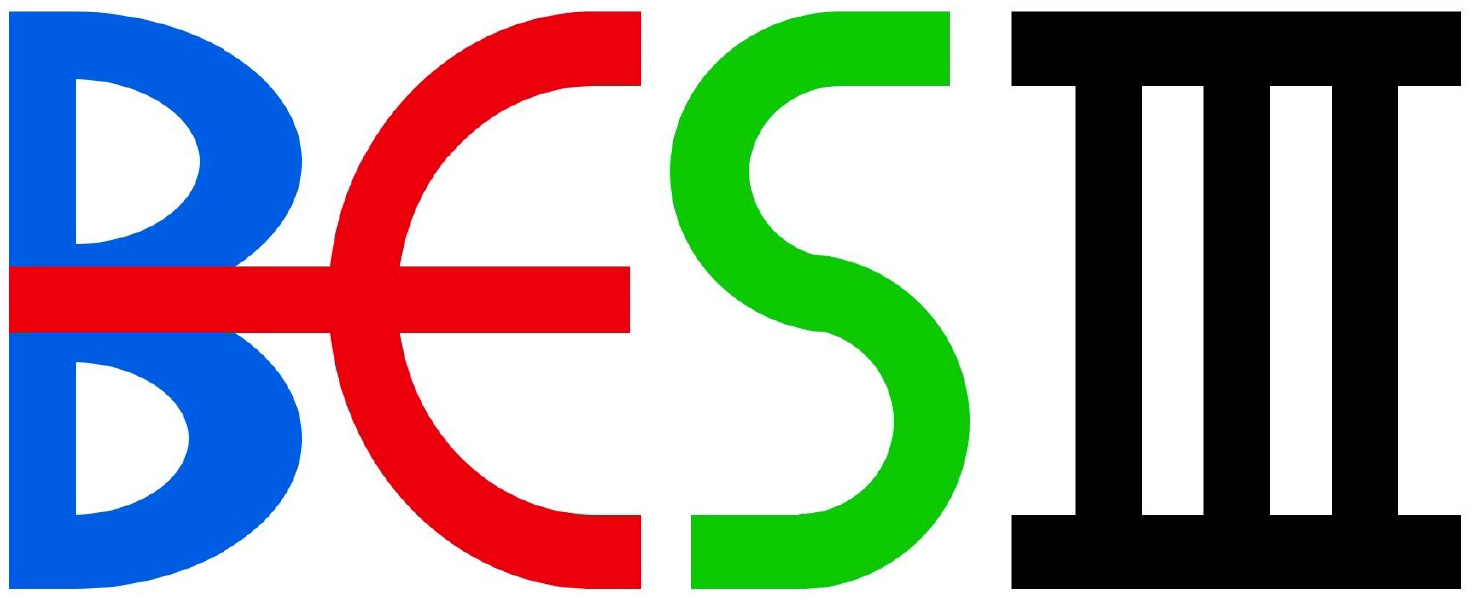}}
\collaboration{The BESIII Collaboration}

\date{\today}
\abstract{
We present the first amplitude analysis of the hadronic decay $D^+\to\pi^+\pi^0\pi^0$, using $e^{+}e^{-}$ collision data collected with the BESIII detector at a center-of-mass energy of 3.773~GeV, corresponding to an integrated luminosity of 20.3~fb$^{-1}$. The fit fractions of the intermediate processes are measured, in which the  $D^+ \to \rho(770)^+\pi^0$ component is found to be dominant with a branching fraction of $(3.08\kern0.15em\pm\kern0.15em0.10_{\rm stat.}\pm0.07_{\rm syst.})\times10^{-3}$. Based on the amplitude analysis, the branching fraction of $D^+ \to \pi^+\pi^0\pi^0$ is measured to be $(4.84\kern0.1em\pm\kern0.1em0.05_{\rm stat.}\kern0.1em\pm\kern0.1em0.05_{\rm syst.})\times10^{-3}$. In addition, the \emph{CP} asymmetries, both for specific amplitudes and integrated over the entire phase space, are measured.
}

\arxivnumber{}
\begin{document}
\newcommand{\BESIIIorcid}[1]{\href{https://orcid.org/#1}{\hspace*{0.1em}\raisebox{-0.45ex}{\includegraphics[width=1em]{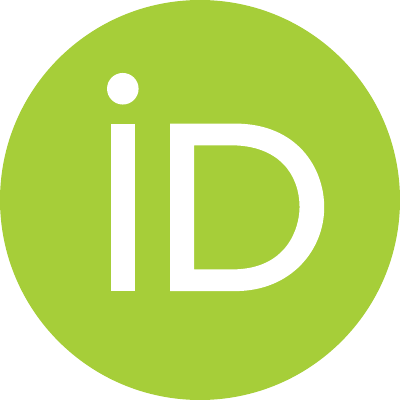}}}}
\maketitle
\flushbottom

\section{Introduction}
\setlength{\parskip}{0pt}

\begin{figure}[htbp]
  \centering
  \includegraphics[width=7cm]{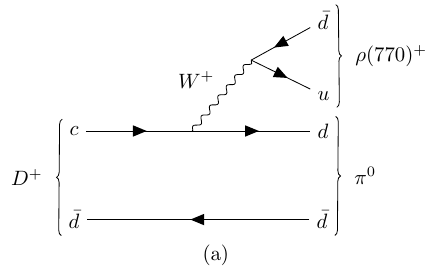}
 \includegraphics[width=7cm]{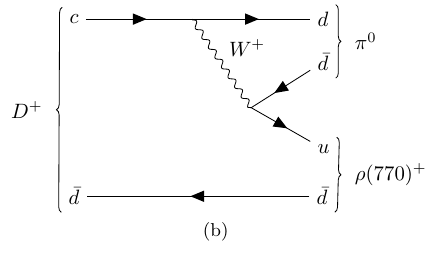}\\
\includegraphics[width=7cm]{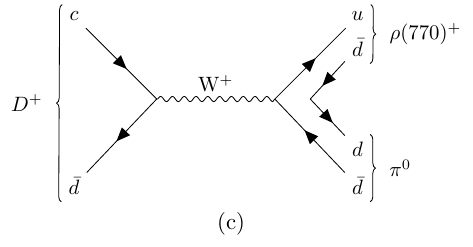}
\includegraphics[width=7cm]{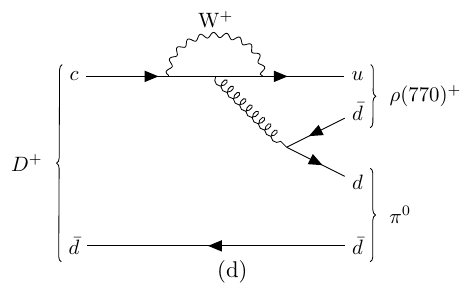}\\
\includegraphics[width=7cm]{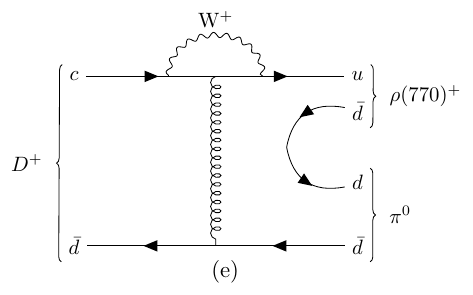}\\
\caption{
Topological diagrams contributing to the decay $D^+\to \rho(770)^+\pi^0$: 
(a) color-favoured external $W$-emission tree diagram, 
(b) color-suppressed internal $W$-emission tree diagram, 
(c) $W$-annihilation diagram, 
 (d) QCD penguin diagram, and 
(e) QCD penguin exchange diagram.
}
  \label{FMT}
\end{figure}

Non-leptonic decays of charmed mesons, characterized by complicated strong interactions and rich topological amplitude structures, provide an ideal platform for in-depth studies of non-perturbative effects in Quantum Chromodynamics (QCD) and tests of weak interaction mechanisms~\cite{PDG,HFLAV}. The three-body decay $D^+ \to \pi^+\pi^0\pi^0$ is expected to be dominated by the quasi-two-body channel $D^+\to\rho(770)^+\pi^0$ (with $\rho(770)^+\to\pi^+\pi^0$). This process is illustrated by the topological diagram shown in Fig.~\ref{FMT}. This channel clearly exhibits the weak decay topology dominated by the color-favoured tree diagram, with possible additional contributions from color-suppressed internal emission, W-annihilation and QCD penguin-type diagrams~\cite{Cheng}.

Theoretical predictions for the branching fraction (BF) of the $D^+\to\rho(770)^+\pi^0$ decay have been obtained within several theoretical frameworks, including the pole model, the topological diagrammatic approaches (TDA) and the factorization-assisted topological-amplitude (FAT) method~\cite{Qin:2014,Cheng:Front2015}. The BF of the $D^+\to\rho(770)^+\pi^0$ decay from different predictions are listed in Table~\ref{tab:theory}. Precise measurements of the BF and the amplitude $D^+\to \pi^+\pi^0\pi^0$ can provide stringent constraints on theoretical models, thus deepening the understanding of the topological amplitude composition~\cite{Yu:2011}.

\begin{table}[htbp]

  \centering
   \caption{Theoretical predictions for the BF of the $D^+\to\rho(770)^+\pi^0$ decay from different theoretical models.}
  \begin{tabular}{lc}
    \hline
    \hline
    Model  &$\mathcal{B}(D^+ \to \rho(770)^+\pi^0)$~($\times 10^{-3}$)\\
    \hline	
    Pole~\cite{Yu:2011}
    &3.5$\pm$1.6 \\
    FAT~[mix]~\cite{Qin:2014}
    &2.5 \\
    TDA~\cite{Cheng:2021}
    &4.44$\pm$0.59 \\
    \hline
    \hline
  \end{tabular}

  \label{tab:theory}
\end{table}

The \emph{CP} violation in charm decays occurs through the interference between processes at tree-level (Figs.~\ref{FMT}(a) and~\ref{FMT}(b)) and processes at higher order (Figs.~\ref{FMT}(d) and~\ref{FMT}(e)) in the singly Cabibbo-suppressed quark transitions $c \to u \bar{d} d$. Within the Standard Model, \emph{CP} asymmetries in charm decays for singly Cabibbo-suppressed processes are predicted to be at the level of $10^{-4}$ to $10^{-3}$~\cite{Grossman:2007}. In 2019, the LHCb experiment reported the first observation of direct \emph{CP} violation in neutral $D^0$ meson decays, with an asymmetry of approximately 0.15\% (on the order of $10^{-3}$), which brought the experimental sensitivity into the predicted theoretical range~\cite{LHCb:2019}. However, no clear \emph{CP} asymmetry has yet been observed in $D^+$ meson decays. References~\cite{xxx,yyy} predict \emph{CP} asymmetries in $D\to \rho\pi$ decays in the range of $(0.3\sim 5.0)\times10^{-4}$. Therefore, searching for \emph{CP} violation in the singly Cabibbo-suppressed decay $D^+\to\pi^+\pi^0\pi^0$ is an interesting pursuit. In addition, \emph{CP} violation in multibody decays can also be induced through long-distance interference effects among different intermediate resonant and non-resonant amplitudes, which may give rise to local \emph{CP} asymmetries across the phase space.

Previously, the BESIII experiment reported the first measurement of \emph{CP} asymmetry in the $D^+\to\pi^+\pi^0\pi^0$ decay based on 2.93~fb$^{-1}$ of data, yielding a \emph{CP} asymmetry with an approximately $2\sigma$ deviation from zero~\cite{BESIII:2022}. Although this result is not statistically significant, it motivates further studies with a much larger data sample. Furthermore, amplitude analysis of three-body $D$ decays can detect \emph{CP} asymmetries in sub-modes within Dalitz plot regions, potentially revealing significantly larger effects in specific intermediate resonance channels, even if the global asymmetry is small~\cite{Cheng:2021}. Therefore, the measurements of \emph{CP} asymmetry in the $D^+\to\pi^+\pi^0\pi^0$ decay are of great significance for probing possible new physics effects.

Utilizing 20.3~fb$^{-1}$~\cite{BESIII:2024lbn} collision data collected with the BESIII detector at a center-of-mass energy of 3.773~GeV, we present the first amplitude analysis and BF measurement of the $D^+ \rightarrow \pi^+\pi^0\pi^0$ decay. A measurement of  the \emph{CP} asymmetry in the decay $D^\pm \rightarrow \pi^\pm\pi^0\pi^0$ is also made. Charge-conjugate modes are implied throughout this paper except when discussing \emph{CP} asymmetries.

\section{Detector and data}
\label{sec:detector_dataset}
The BESIII detector~\cite{ABLIKIM2010345} records symmetric $e^+e^-$ collisions provided by the BEPCII storage ring~\cite{Yu:2016cof} in the center-of-mass energy range from 1.84 to 4.95~GeV, achieving a peak luminosity of $1.1 \times 10^{33}\;\text{cm}^{-2}\text{s}^{-1}$ at $\sqrt{s} = 3.773\;\text{GeV}$. The BESIII has collected large data samples in this energy region~\cite{Ablikim_2020,4444,5555}. The cylindrical core of the BESIII detector covers 93\% of the full solid angle and comprises a helium-based multilayer drift chamber~(MDC), a plastic scintillator time-of-flight system~(TOF), and a CsI(Tl) electromagnetic calorimeter~(EMC), all enclosed in a superconducting solenoidal magnet providing a 1.0~T magnetic field. The solenoid is supported by an octagonal flux-return yoke with resistive plate counter muon identification modules interleaved with steel. The charged-particle momentum resolution at $1~{\rm GeV}/c$ is $0.5\%$, and the ${\rm d}E/{\rm d}x$ resolution is $6\%$ for electrons from Bhabha scattering. The EMC measures photon energies with a resolution of $2.5\%$ ($5\%$) at $1$~GeV in the barrel (end cap) region. The time resolution in the TOF barrel region is 68~ps, while that in the end cap region was 110~ps. The end cap TOF system was upgraded in 2015 using multigap resistive plate chamber technology, providing a time resolution of 60~ps, benefiting 85.6\% of the data used in this analysis~\cite{TOF}.

Simulated inclusive Monte Carlo (MC) samples are produced with a {\sc geant4}-based~\cite{geant4} MC simulation package, which includes the geometric description of the BESIII detector~\cite{NST33142} and the detector response. These simulations are used to determine detection efficiencies and to estimate backgrounds. The simulation models the beam energy spread and initial state radiation~(ISR) in the $e^+e^-$ annihilations with the generator {\sc kkmc}~\cite{kkmc}. The inclusive MC samples consist of the production of $D\bar{D}$ pairs, the non-$D\bar{D}$ decays of the $\psi(3770)$, the ISR production of the $J/\psi$ and $\psi(3686)$ states, and the continuum processes incorporated in {\sc kkmc}. All particle decays are modeled with {\sc evtgen}~\cite{evtgen} using BF either taken from the Particle Data Group (PDG)~\cite{PDG} when available or estimated with {\sc lundcharm}~\cite{lundcharm}. Final state radiation from charged final state particles is incorporated using {\sc photos}~\cite{photos}. The phase-space (PHSP) MC sample is generated with a uniform distribution for the decay $D^+ \rightarrow \pi^+ \pi^0 \pi^0$, which is used to calculate the normalization integral of the probability density function (PDF) in the amplitude analysis. The signal MC sample, which is used to determine the detection efficiency in the BF measurement, is generated according to the amplitude analysis results.

\section{Event selection}
\label{ST-selection}
The process $e^{+}e^{-} \to \psi(3770) \to D^{+}D^{-}$ allows the studies of $D$ decays with a tag technique~\cite{MarkIII-tag}. There are two types of samples used in the tag technique: single tag~(ST) and double tag~(DT). In the ST sample, only one $D^{-}$ meson is reconstructed through a particular hadronic decay mode. In the DT sample, the signal $D^{+}$ meson is reconstructed through the $D^{+} \to \pi^{+}\pi^{0}\pi^{0}$ decay, while the tag $D^{-}$ meson is reconstructed through one of the six hadronic decay modes: $D^-\to K^{+}\pi^{-}\pi^{-}$, $K^{+}\pi^{-}\pi^{-}\pi^{0}$, $K^{0}_{S}\pi^{-}$, $K_{S}^{0}\pi^{-}\pi^{0}$, $K_{S}^{0}\pi^{-}\pi^{-}\pi^{+}$, and $K^{+}K^{-}\pi^{-}$. The $D^\pm$ candidates are constructed from  individual $K^\pm$, $\pi^\pm$, $K_S^0$ and $\pi^0$ candidates with the following selection criteria. 

Charged tracks detected in the MDC must satisfy $|$cos$\theta|<0.93$, where $\theta$ is defined as the polar angle with respect to the $z$-axis, which is the symmetry axis of the MDC. For the charged tracks not originating from $K_S^0$ decays, the distance of the closest approach to the interaction point (IP) is required to be less than 10\,cm along the $z$-axis, $|V_{z}|$, and less than 1\,cm in the transverse plane, $|V_{xy}|$. Particle identification (PID) for charged tracks combines the measurements of the ${\rm d}E/{\rm d}x$ in the MDC and the flight time in the TOF to form probabilities $\mathcal{L}(h)~(h=K,\pi)$ for each hadron ($h$) hypothesis. The charged kaons (pions) are identified by comparing the likelihoods for the kaon and pion hypotheses, and are selected if $\mathcal{L}(K)>\mathcal{L}(\pi)$ ($\mathcal{L}(\pi)>\mathcal{L}(K)$). 

Each $K_{S}^0$ candidate is reconstructed from two oppositely charged tracks satisfying $|V_{z}|<$ 20 cm. The two charged tracks are assigned as $\pi^+\pi^-$ without imposing further PID criteria. Fits to the primary and secondary vertices are performed, and the decay length of the $K_{S}^0$ candidate is required to be greater than twice the vertex resolution away from the IP. The $\chi^2$ of the vertex fit must be less than 100 and the invariant mass of the $\pi^{+} \pi^{-}$ pair $\left(M_{\pi^{+} \pi^{-}}\right)$ is required to be in the range $[0.487,0.511]~\mathrm{GeV} / c^2$ to form the candidate $K_S^0$ particles.

Photon candidates are identified using isolated showers in the EMC. The deposited energy of each shower must be more than 25~MeV in the barrel region ($|\cos \theta|< 0.80$) and more than 50~MeV in the end cap region ($0.86 <|\cos \theta|< 0.92$). To exclude showers that originate from charged tracks, the angle subtended by the EMC shower and the position of the closest charged track at the EMC must be greater than 10 degrees as measured from the IP. To suppress electronic noise and showers unrelated to the event, the difference between the EMC time and the event start time is required to be within 
[0, 700]\,ns.

The $\pi^0$ candidates are reconstructed from the photon pairs with invariant masses in a range of \mbox{[0.115, 0.150]} $\mathrm{GeV} / c^2$, which is about five times the mass resolution. Moreover, in order to achieve an adequate resolution, at least one of the two photons is required to be detected in the barrel EMC. A one-constraint kinematic fit that constrains the $\gamma \gamma$ invariant mass to the known $\pi^0$ mass~\cite{PDG} is performed to improve the mass resolution. The $\chi^2$ of the kinematic fit is required to be less than 50 .

To distinguish the $D^\pm$ mesons from combinatorial backgrounds, we define two variables, the beam constrained mass $M_{\mathrm{BC}}$ and the energy difference $\Delta E$
$$
M_{\mathrm{BC}}=\sqrt{E_{\mathrm{beam}}^2-\left|\vec{p}_{D^\pm}\right|^2},~\Delta E=E_{D^\pm}-E_{\mathrm{beam}},
$$where $\vec{p}_{D^\pm}$ and $E_{D^\pm}$ are the total reconstructed momentum and energy of the $D^\pm$ candidate, and $E_{\text {beam }}$ is the beam energy.

The $D^\pm$ signals  appear as a peak at the known $D^\pm$ mass~\cite{PDG} in the $M_{\rm BC}$ distribution and as a peak at zero in the $\Delta{E}$ distribution. For both signal and tag modes, candidates are required to satisfy $1.865<M_{\rm BC}<1.875$~GeV/$c^2$. The $\Delta E$ windows are listed in Table~\ref{tagdeltaE}. 

\begin{table}[htp]
	\centering
\caption{Requirements of $\Delta E$ for signal mode and different $D^{-}$ tag modes. }
	\begin{tabular}{llllll ccc}
		\hline
		\hline
		&  Decay mode  &$\Delta E$(GeV)   \\
		\hline
        &  $D^+\to\pi^+\pi^0\pi^0$ &  $(-0.100,~ 0.045)$ \\
		\hline
        &  $D^{-} \to K^{+}\pi^{-}\pi^{-}$    &  $(-0.025,~ 0.024)$ \\
		&  $D^{-}\to K^{+}\pi^{-}\pi^{-}\pi^{0}$ &$(-0.057,~ 0.046)$ \\
		&  $D^{-}\to K_{S}^{0}\pi^{-}$ & $(-0.025,~ 0.026)$ \\
		&  $D^{-}\to K_{S}^{0}\pi^{-}\pi^{0}$& $(-0.062,~ 0.049)$ \\
		&  $D^{-}\to K_{S}^{0}\pi^{-}\pi^{-}\pi^{+}$ & $(-0.028,~ 0.027)$\\ 
		& $D^{-}\to K^{+}K^{-}\pi^{-}$ & $(-0.024,~ 0.023)$\\
		\hline
		\hline
	\end{tabular}
	\label{tagdeltaE}
\end{table}

\section{Amplitude analysis}
\label{Amplitude-Analysis}
\subsection{Further selection criteria in amplitude analysis}
\label{AASelection}
To suppress the background coming from $D^+\to K_S^0\pi^+$ $(K_S^0 \to \pi^0\pi^0)$, a $K_S^0$ veto is applied. Events with $0.428<M_{\pi^0\pi^0}<0.548$ $\mathrm{GeV} / c^2$ are discarded, where $M_{\pi^0\pi^0}$ denotes the invariant mass of the $\pi^0\pi^0$  system. The $D^0\to K^-\pi^+\pi^0$ versus $\bar{D^0}\to K^+\pi^-\pi^0$ decay process can fake the signal process $D^+\to \pi^+\pi^0\pi^0$ versus $D^- \to K^+K^-\pi^-$ through the exchange of a $K^-$ meson from $D^0$ and a $\pi^0$ meson from $\bar{D^0}$, and is referred to as mispartition background. To suppress it we reject the events that simultaneously
satisfy $\left|M_{K^{-} \pi^{+} \pi^0}-1.865\right|<0.05~ \mathrm{GeV} / c^2$ and $\left|M_{K^{+} \pi^{-} \pi^0}-1.865\right|<0.05~ \mathrm{GeV} / c^2$, where, $M_{K^-\pi^+\pi^0}$ and $M_{K^+\pi^-\pi^0}$ denote the invariant masses of the $K^-\pi^+\pi^0$ and $K^+\pi^-\pi^0$ systems, respectively.

If there exist multiple signal candidates in an event, the candidate with minimum value of $\Delta E_{\text {tag }}^2$+$\Delta E_{\text {sig }}^2$ is retained. To optimize resolution and  ensure events lie within the PHSP boundary, a seven-constraint kinematic fit is performed. In this fit, the four-momenta of the final-state particles are constrained to the initial four-momenta of the $e^{+}e^{-}$ system, and the reconstructed masses of the $D^+$ and the two $\pi^0$ candidates are constrained to their known values~\cite{PDG}. The modified four-momenta of the final-state particles from the kinematic fit are used in the amplitude analysis.

\begin{figure}[htp]
  \centering
    \includegraphics[width=0.6\textwidth]{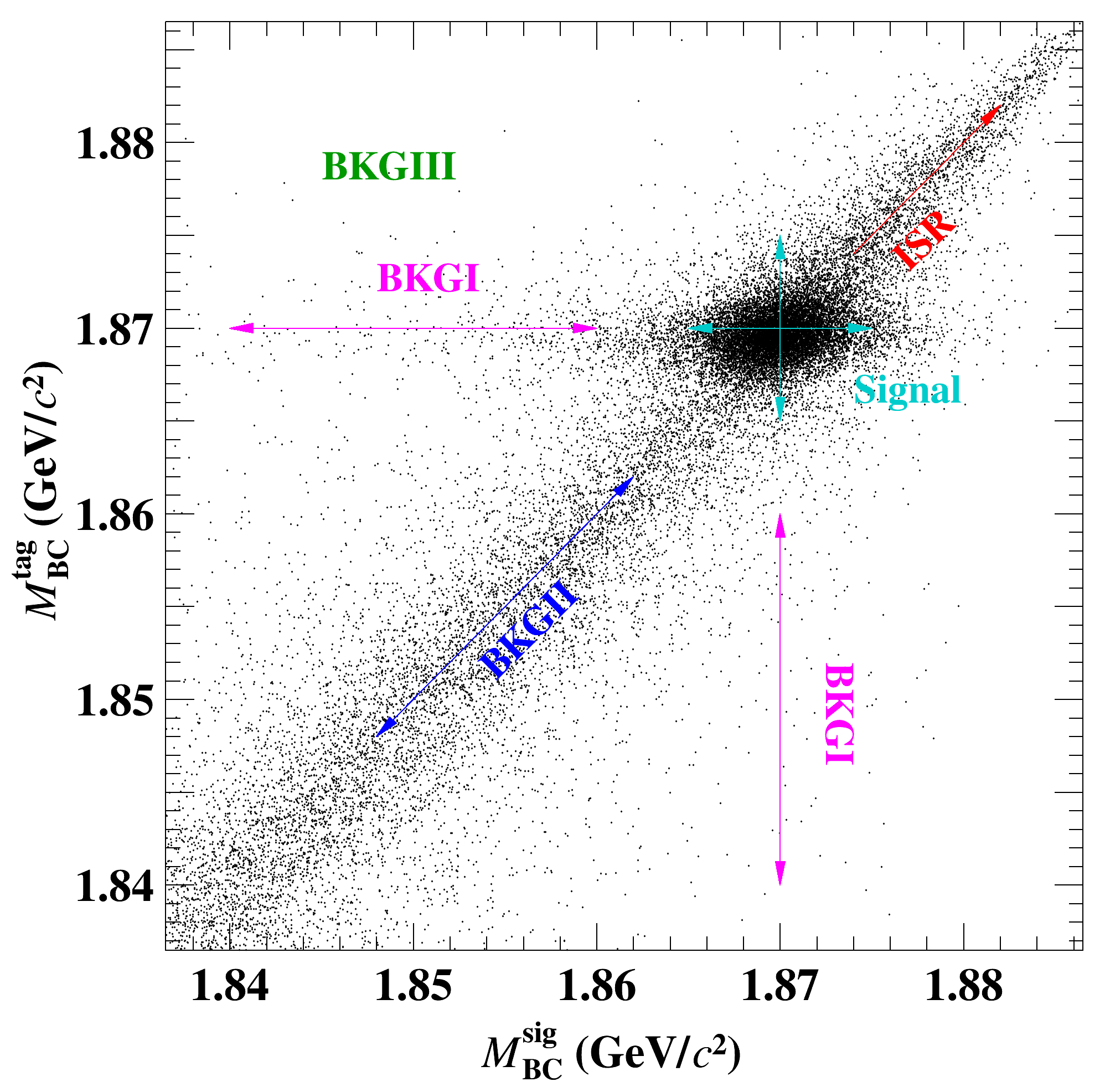}
  \caption{The 2D distributions of $M_{\rm BC}^{\text{tag}}$ versus $M_{\rm BC}^{\text{sig}}$ for the $D^+ \to \pi^+ \pi^0 \pi^0$ candidates selected from the inclusive MC sample, illustrating the different components. The cyan arrows indicate the signal region, the red arrow denotes the ISR effect. The pink and blue arrows indicate the background due to incorrectly reconstructed $D^-$ or $D^+$ and mispartitioning continuum, respectively.}
  \label{2DMBC}
\end{figure}

An unbinned two-dimensional (2D) maximum likelihood fit is performed on the distribution of $M_{\text{BC}}^{\text{tag}}$ versus $M_{\text{BC}}^{\text{sig}}$ to extract the signal yields, where $M_{\rm BC}^{\text{sig}}$ and $M_{\rm BC}^{\text{tag}}$ denote the beam-constrained masses of the signal-side and tag-side candidates, respectively. Signal events with both tag and signal sides reconstructed correctly, will concentrate around $M_{\text{BC}}^{\text{sig}} = M_{\text{BC}}^{\text{tag}} = M_D$, where $M_D$ is the nominal $D$ mass. We define three kinds of background events.  Candidates with correctly reconstructed $D^+$(or $D^-$) and incorrectly reconstructed $D^-$(or $D^+$) are marked as BKGI, which appear around the lines $M_{\text{BC}}^{\text{sig}}$ or $M_{\text{BC}}^{\text{tag}} = M_D$. Other candidates that appear around the diagonal are mainly from the mispartition of events containing a $D^0\bar{D}^0$ pair and the $e^+e^-\to q\bar{q}$ processes (BKGII). The remaining flat backgrounds mainly come from candidates reconstructed incorrectly on both sides (BKGIII). The PDFs for the different components used in the fit are the following:
\begin{itemize}
	\item Signal: $s(x,y)$,
	\item BKGI: $b_1(x)\cdot A(y;m_0,c,p)+b_2(y)\cdot A(x;m_0,c,p)$,
	\item BKGII: $A((x+y)/\sqrt{2};m_0,c,p)\cdot g((x-y)/\sqrt{2})$,
	\item BKGIII: $A(x;m_0,c,p)\cdot A(y;m_0,c,p)$,
\end{itemize}
where $x$, $y$ are the $M_{\text{BC}}^{\text{sig}}$ and $M_{\text{BC}}^{\text{tag}}$, respectively.
The signal shape $s(x, y)$ is described by the 2D MC-simulated shape convolved with a 2D Gaussian. The Gaussian function is included to correct for the resolution differences between data and MC. The parameters of the Gaussian function are obtained by 1D fits to the $M_{\rm BC}$ distributions on the signal and tag sides, respectively, and are fixed in the 2D fit. For BKGI, $b_{1}(x)$ or $b_{2}(y)$ are the one-dimensional MC-simulated shapes convolved with a Gaussian, and $\operatorname{A}(z)$ is an ARGUS function~\cite{Albrecht:1990am} in the opposing coordinate. For BKGII, $\operatorname{A}((x+y)/\sqrt{2})$ is an ARGUS function in the diagonal axis, and is multiplied by $g((x-y)/\sqrt{2})$ which is a Gaussian in the perpendicular axis. The PDF for BKGIII is an ARGUS in both directions multiplied together. Fig.~\ref{2DMBC} shows the 2D distributions of $M_{\rm BC}^{\text{tag}}$ versus $M_{\rm BC}^{\text{sig}}$ for the $D^+ \to \pi^+ \pi^0 \pi^0$ candidates selected from the inclusive MC sample, illustrating the different components. In the fit, the parameters $m_0$ and $p$ for the ARGUS function are fixed at $1.8865~\mathrm{GeV} / c^2$ and 0.5, respectively.

\begin{figure}[htbp]
  \centering
\begin{minipage}[t]{0.49\textwidth}
  \centering
  \includegraphics[width=\textwidth]{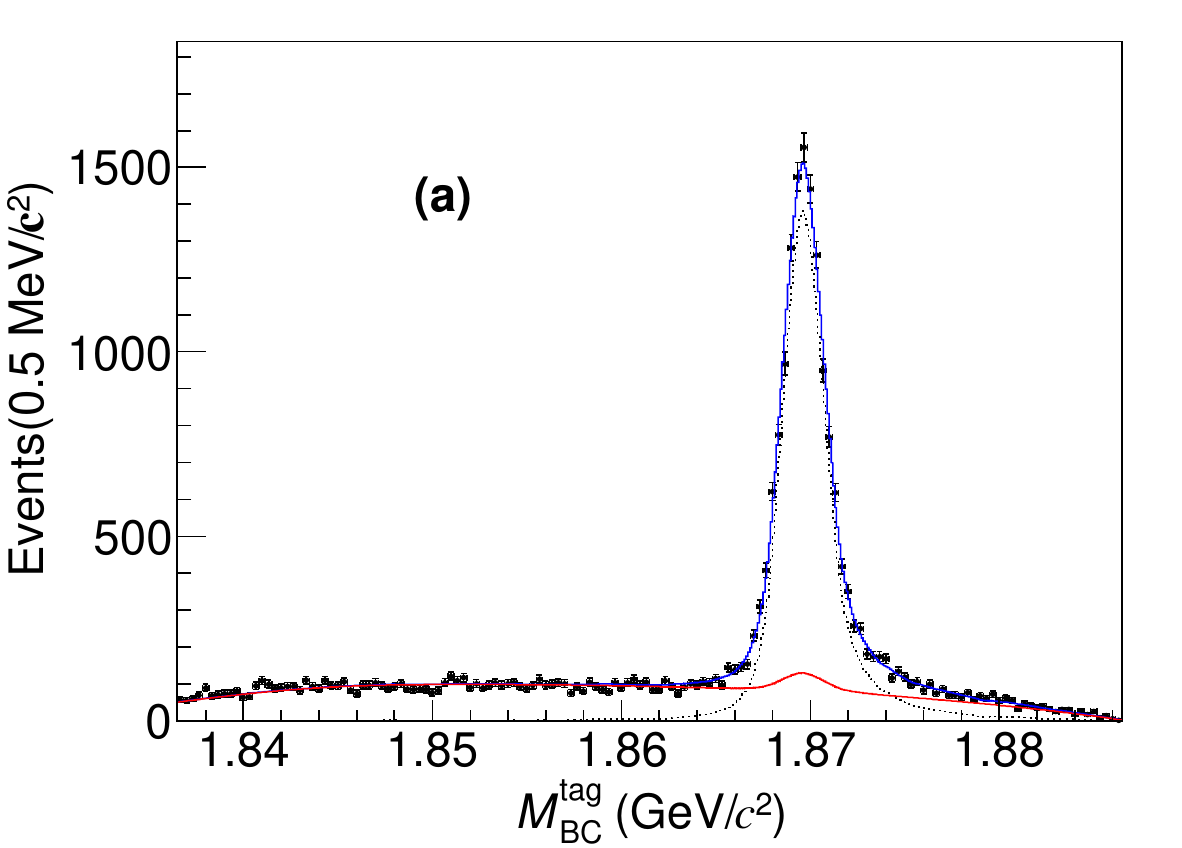}
\end{minipage}
\hfill
\begin{minipage}[t]{0.49\textwidth}
  \centering
  \includegraphics[width=\textwidth]{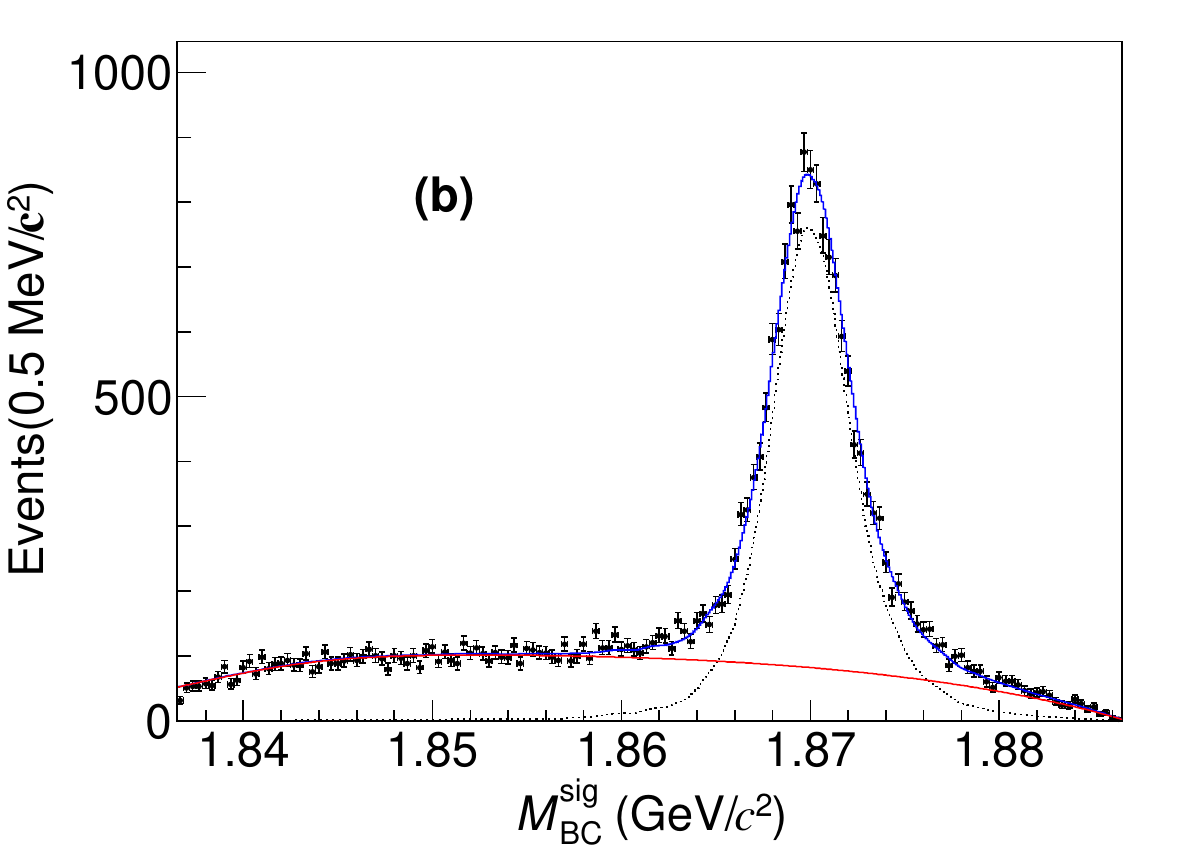}
\end{minipage}
  \caption{Projections on $M_{\mathrm{BC}}^{\mathrm{tag}}$ (a) and $M_{\mathrm{BC}}^{\text {sig }}$ (b) of the 2D fit. The points with error bars are data. The blue line is the total fit result, the dotted line is the signal shape and the red line is the combinatorial background.}
  \label{BF2D}
\end{figure}

 The signal region is defined as  $1.865<M_{\text{BC}}^{\text{tag}}<1.875$~GeV/$c^2$, $1.865<M_{\text{BC}}^{\text{sig}}<1.875$~GeV/$c^2$, and the retained data sample in the signal region contains 13498 events with a signal purity of $(87.4 \pm 0.3)\%$. The projections of the 2D fit on data are shown in Fig.~\ref{BF2D}.

\subsection{Fit method}
The amplitude analysis of $D^+ \to \pi^+ \pi^0 \pi^0$ is performed by an unbinned maximum likelihood fit. The likelihood function $\mathcal{L}$ is constructed with a combined signal-background probability density function (PDF). The log-likelihood is written as
\begin{equation}
  \ln{\mathcal{L}} = \begin{matrix} \sum\limits_{k}^{N_{D}} \ln [\omega f_{S}(p^{k})\end{matrix}+(1-\omega)f_B(p^k)],  \label{loglikelihood}
\end{equation}
where $p^k$ denote the four-momenta of the final state particles, and the index $k$ denotes the $k^{\rm th}$ event in data. The number of candidates in data is given by $N_D$, $f_S(f_B)$ is the signal (background) PDF and $\omega$ is the purity of the signals discussed in Sec.~\ref{AASelection}.

The signal PDF $\it{f_S(p)}$ is given by 
\begin{equation}
\begin{aligned}
    \it{f_S}(p) = \frac{\epsilon(p)|\mathcal{M}(p)|^{\rm{2}}R_{\rm{3}}(p)}{\int \epsilon(p)|\mathcal{M}(p)|^{\rm{2}}R_{\rm{3}}(p)dp},
\label{pwa:pdf}
\end{aligned}
\end{equation}
where $\epsilon(p)$ is the detection efficiency and $R_3(p)$ is the three-body PHSP factor. $R_3(p)$ is defined as
\begin{equation}
R_3(p) = \delta\left(p_{D^+}-\sum\limits_{j=1}^{3}p_j\right)\prod\limits_{j=1}^{3}\delta\left(p_j^2-m^2\right) \; \theta(E_j),
\label{pwa:R3}
\end{equation}
where $j$ runs over the three decay products, $E_j$ is the energy of particle $j$, and $\theta(E_j)$ is the step function. The total amplitude $\mathcal{M}$ is treated with the isobar model~\cite{Fleming1964Isobar},
which uses the coherent sum of the amplitudes of the intermediate processes, $\mathcal M(p) = \sum{c_n\mathcal A_n(p)}$,
where the $c_n = \rho_ne^{i\phi_n}$ are complex coefficients. The magnitude $\rho_n$ and phase $\phi_n$ are the free parameters in the fit.
The amplitude of the $n^{\rm th}$ intermediate state ($\mathcal A_n$) is
\begin{equation}
	\mathcal A_n(p) = P_n(p)S_n(p)F^r_n(p)F^D_n(p),
\end{equation}
where $P_n(p)$ is the propagator of the intermediate resonance, $S_n(p)$ is the spin factor, $F^{r}_n(p)$ and $F^{D}_n(p)$ are the Blatt-Weisskopf barrier factors for the intermediate resonance and $D^+$, respectively. Since the final state contains two identical $\pi^0$ mesons, the total amplitude is Bose symmetrized by summing over the exchange of the two $\pi^0$ mesons at the amplitude level before taking the modulus squared.

The background PDF is given by
\begin{equation}
f_B(p) = \frac{B(p)R_{3}(p)}{\int B(p)R_{3}(p)dp},
\end{equation}
where $B(p)$ represents the background function. In the numerator of Eq.~(\ref{pwa:pdf}), the $\epsilon(p)$ and $R_{3}(p)$ terms which are independent of the fitted variables, are regarded as constants and can be dropped during the fit from the log-likelihood, Eq.~(\ref{loglikelihood}). To extract the shared component $\epsilon(p)R_{3}(p)$, the background PDF can be expressed as follows:
\begin{equation}
f_B(p) = \frac{\epsilon(p)B_{\epsilon}(p)R_{3}(p)}{\int \epsilon(p)B_{\epsilon}(p)R_{3}(p)dp},
\end{equation}
where $B_{\epsilon}(p)=B(p)/\epsilon(p)$ is the efficiency-corrected background shape. 
The shape of the background in data is modeled by the background events in the signal region derived from the inclusive MC samples. 
The validity of this description is checked by comparing the 1D projections of $M_{\pi^+\pi^0}$ and $M_{\pi^0\pi^0}$ of events outside the $M_{\mathrm{BC}}$ signal region between data and MC. 
The distributions of the background events in the inclusive MC samples, both inside and outside the $M_{\mathrm{BC}}$ signal region, have also been examined. Generally, they are compatible with each other within statistical uncertainties. The selected inclusive MC samples in the signal region, after excluding the signal events, are collected, and their invariant masses ($M_{\pi^+\pi^0}$ and $M_{\pi^0\pi^0}$) are used as input for the XGBoost package~\cite{XGboost}, in which the two $\pi^0$s are Bose symmetrized at the PDF level. By training the inclusive MC samples and PHSP MC samples using different XGBoost models, we obtain the background probability $B(p) R_3(p)$ and the efficiency probability $\epsilon(p) R_3(p)$, respectively. Dividing the two probabilities yields the value of $B_\epsilon(p)$.

As a consequence, the log-likelihood can be written as
\begin{equation}
  \ln{\mathcal{L}} =  \sum\limits_{k=1}^{N_{D}}{\rm ln}\Bigg[ \omega \frac{|\mathcal{M}(p^k)|^2}{\int \epsilon(p^k)|\mathcal{M}(p^k)|^2R_3(p^k)dp^k}+(1-\omega)\frac{B_{\epsilon}(p^k)}{\int{\epsilon(p^k) B_\epsilon(p^k)R_3(p^k)}dp^k}\Bigg].
  \label{likelihoodfinal}
\end{equation}
The normalization integrals of signal and background are evaluated by MC integration,
\begin{equation}
\begin{aligned}
&\int \epsilon(p)|\mathcal{M}(p)|^2R_3(p)\,dp \propto \frac{1}{N_{\rm MC}}\sum^{N_{\rm MC}}_{k=1}\frac{|\mathcal{M}(p^{k})|^2}{|\mathcal{M}^{\rm gen}(p^{k})|^2},\\
&\int \epsilon(p)B_\epsilon(p)R_3(p)dp \propto \frac{1}{N_{\rm MC}}\sum^{N_{\rm MC}}_{k=1}\frac{B_\epsilon(p^{k})}{|\mathcal{M}^{\rm gen}(p^{k})|^2},
\end{aligned}
\end{equation}
where $N_{\rm MC}$ is the number of the selected MC events. 
The $M^{\rm gen}(p)$ is the signal PDF used to generate the MC samples in MC integration.

Differences in tracking, PID and $\pi^0$ reconstruction efficiencies between data and MC simulation are corrected by weighting each MC event with a factor $\gamma_{\epsilon}$, defined as
\begin{equation}
  \gamma_{\epsilon}(p) = \prod_{n} \frac{\epsilon_{n,\rm data}(p)}{\epsilon_{n,\rm MC}(p)},
  \label{pwa:gamma}
\end{equation}
where $n$ refers to tracking, PID and $\pi^0$ reconstruction,  $\epsilon_{n,\rm data}(p)$ and $\epsilon_{n,\rm MC}(p)$ are the tracking, PID and $\pi^0$ reconstruction efficiencies as a function of the decay-product momenta in data and MC simulation, respectively.
By weighting each signal MC event with $\gamma_{\epsilon}$, the MC integration is given by
\begin{equation}
  \int \epsilon(p)|\mathcal M(p)|^2R_3(p)dp \propto \frac{1}{N_{\rm MC}}\sum^{N_{\rm MC}}_{k=1}\frac{\gamma_{\epsilon}(p^{k_{\rm MC}})|\mathcal{M}(p^{k})|^2}{|\mathcal{M}^{\rm gen}(p^{k})|^2}.\label{likelihood3}
\end{equation}

\subsubsection{Blatt-Weisskopf barriers}
\label{Blatt-Weisskopfbarriers}
For a decay process $a \to b+c$, the Blatt-Weisskopf barriers~\cite{PhysRevD.104.012016} depend on the angular momenta $L$ 
and the momentum $q$ of the final-state particle $b$ or $c$ in the rest system of $a$. 
They are taken as
\begin{equation}
\begin{aligned}
  &X_{L=0}(q)=1,\\
  &X_{L=1}(q)=\sqrt{\frac{z_0^2+1}{z^2+1}},\\
  &X_{L=2}(q)=\sqrt{\frac{z_0^4+3z_0^2+9}{z^4+3z^2+9}}, \label{xl}
\end{aligned}
\end{equation}
where $z = qR$ and $z_0 = q_0R$ with $q_0$ defined in Sec.~\ref{Propagator}. The effective radius of barrier $R$ is fixed to be 3.0 $(\mathrm{GeV} /c)^{-1}$ for the intermediate resonances and 5.0 $(\mathrm{GeV} /c)^{-1}$ for the $D^+$ meson~\cite{RRR}.

\subsubsection{Propagator}
\label{Propagator}

The intermediate resonance $f_2(1270)$ is parameterized with the relativistic Breit-Wigner (RBW) formulas~\cite{Jackson:1964res},
\begin{equation}
\begin{aligned}
		&P(m) = \frac{1}{(m^2_0-m^2)-im_0\Gamma(m)}, \\ 
		&\Gamma(m)=\Gamma_0\left(\frac{q}{q_0}\right)^{2L+1}\Big(\frac{m_0}{m}\Big)\left(\frac{X_L(q)}{X_L(q_0)}\right)^2,  \label{propagator}
\end{aligned}
\end{equation}

where $m^2$ is the invariant mass squared of the decay products of the intermediate resonance, $m_0$ and $\Gamma_0$ are the mass and width of the intermediate resonance, which are fixed to the PDG values for $f_2(1270)$. In a process $a \to b+c$, the variable $q$ is defined as
\begin{equation}
q = \sqrt{\frac{(s_a+s_b-s_c)^2}{4s_a}-s_b}, \label{q2}
\end{equation}

where $s_a, s_b,$ and $s_c$ are the invariant-masses squared of particles $a,\ b$ and $c$, respectively. In Eq.~(\ref{propagator}), we set $q_0 = q(m_0^2)$.

The Gounaris-Sakurai (GS) function~\cite{Gounaris:1968mw}, which incorporates dispersive corrections to the resonance mass, is used to parameterize the spin-1 $\rho$-type resonances $\rho(770)$ and $\rho(1450)$. It serves as a modified RBW lineshape, represented as

\begin{equation}
P_{\rm GS}(m)=\frac{1+d\frac{\Gamma_0}{m_0}}{(m^2_0-m^2)+f(m)-im_0\Gamma(m)},
\end{equation}
where
\begin{equation}
	f(m)=\Gamma_0\frac{m^2_0}{q^3_0}\left[q^2(h(m)-h(m_0))+(m^2_0-m^2)q^2_0\frac{dh}{d(m^2)}\Big|_{m^2=m^2_0}\right]
\end{equation}
and the function $h(m)$ is defined as
\begin{equation}
h(m)=\frac{2}{\pi}\frac{q}{m}{\rm ln}\left(\frac{m+2q}{2m_{\pi}}\right),
\end{equation}
with
\begin{equation}
\frac{dh}{d(m^2)}\Big|_{m^2=m_0^2} = h(m_0)[(8q^2_0)^{-1}-(2m^2_0)^{-1}]+(2\pi m^2_0)^{-1},
\end{equation}
where $m_{\pi}$ is the mass of $\pi$, and the normalization condition at $P_{\rm GS}(0)$ fixes the parameter $d=\frac{f(0)}{\Gamma_0m_0}$. It is found to be
\begin{equation}
d=\frac{3}{\pi}\frac{m^2_{\pi}}{q^2_0}ln\left(\frac{m_0+2q_0}{2m_{\pi}}\right)+\frac{m_0}{2\pi q_0}-\frac{m^2_{\pi}m_0}{\pi q^3_0}.
\end{equation}

The $\pi^0 \pi^0$ $S$-wave is described by the $K$-matrix parametrization. Detailed descriptions of the $K$-matrix formalism can be found in the following references~\cite{K-matrix1,K-matrix2}. The ``$K$-matrix amplitude’’ is referred to as the product of the production vector $P$ and the matrix propagator ($I - iK\rho)^{-1}$
\begin{equation}
A_i=(I-i K \rho)_{i j}^{-1} P_j,
\end{equation}
where $I$ is the identity matrix, and $K$ is the $K$-matrix describing the scattering process and $\rho$ is the PHSP matrix. The indices $i$ and $j$ represent the coupled channel $(1=\pi \pi$, $2=K \bar{K}$, $3=4 \pi$, $4=\eta \eta, 5=\eta \eta^{\prime})$.
The $K$ matrix is expressed as
\begin{equation}
K_{i j}(s)=\left(\sum_\alpha \frac{g_i^\alpha g_j^\alpha}{m_\alpha^2-s}+f_{i j}^{\text {scatt }} \frac{1-s_0^{\text {scatt }}}{s-s_0^{\text {scatt }}}\right)\left[\frac{1-s_{A_0}}{s-s_{A_0}}\left(s-s_A m_\pi^2 / 2\right)\right] \text {, }
\end{equation}
where $g_i^\alpha$ denote the real coupling constants of the pole $m_\alpha$ to the meson channel $i$. The parameters $f_{i j}^{\text {scatt }}$ and $s_0^{\text {scatt }}$ describe a smooth part for the $K$-matrix elements.
The $P$ vector is given by
\begin{equation}
P_j(s)=f_{1 j}^{\text {prod }} \frac{1-s_0^{\text {scatt }}}{s-s_0^{\text {scatt }}}+\sum_\alpha \frac{\beta^\alpha g_j^\alpha}{m_\alpha^2-s},
\end{equation}
where the $\beta_\alpha$ are the complex production couplings, and the parameters $f_{1 j}^{\text {prod }}$ and $s_0^{\text {prod }}$ describe the production of the slowly varying part of the $K$-matrix.

In this analysis, the $K$-matrix parameters $m_\alpha, g_i^\alpha$, $f_{i j}^{\text {scatt }}$, $s_0^{\text {scatt }}, s_{A 0}$, and $s_A$ are fixed to the results in Ref.~\cite{K-matrix1}. The complex production couplings $\beta_\alpha$ and the production parameters $f_{1 j}^{\text {prod }}$ are free parameters determined from the fit.

\subsubsection{Spin factors}
\label{Spinfactors}
The spin-projection operators are defined as~\cite{covariant-tensors}
\begin{equation}
\begin{aligned}
&P^0(a) = 1,&(\mathcal{S}\ \rm wave)\\
&P^{(1)}_{\mu\nu}(a)=-g_{\mu\nu}+\frac{p_{a\mu}p_{a\nu}}{p^2_a},&(\mathcal{P}\ \rm wave)\\
&P^{(2)}_{\mu_1\mu_2\nu_1\nu_2}(a)=\frac{1}{2}\left(P^{(1)}_{\mu_1\nu_1}P^{(1)}_{\mu_2\nu_2}+P^{(1)}_{\mu_1\nu_2}P^{(1)}_{\mu_2\nu_1}\right)-\frac{1}{3}P^{(1)}_{\mu_1\mu_2}P^{(1)}_{\nu_1\nu_2}.&(\mathcal{D}\ \rm wave)
\end{aligned}
\end{equation}
In the two-body decay $a\to b +c$, the $p_a,\ p_b$, and $p_c$ variables are the momenta of particles $a,\ b$, and $c$, respectively, and $r_a = p_b - p_c$. The covariant tensors are given by
\begin{equation}
\begin{aligned}
&\tilde{t}^{(0)}(a) = 1,&(\mathcal{S}\ {\rm wave})\\
&\tilde{t}^{(1)}_\mu(a) = -P^{(1)}_{\mu\nu}(a)r_a^{\nu},&(\mathcal{P}\  \rm wave)\\
&\tilde{t}^{(2)}_{\mu\nu}(a) = P^{(2)}_{\mu\mu_1\nu\nu_2}(a)r_a^{\mu_1}r_a^{\nu_1}.&(\mathcal{D}\ \rm wave)
\end{aligned}
\end{equation}
The spin factors for the $\mathcal{S},\mathcal{P}$, and $\mathcal{D}$ wave decays are
\begin{equation}
\begin{aligned}
&S_n = 1,&(S\ \rm wave)\\
&S_n = \tilde{T}^{(1)\mu}(D)\tilde{t}^{(1)}_\mu(a),&(P\ \rm wave)\\
&S_n = \tilde{T}^{(2)\mu\nu}(D)\tilde{t}^{(2)}_{\mu\nu}(a),&(D\ \rm wave)
\end{aligned}
\end{equation}
where $\tilde{T}$ is the tensor describing the $D^{+}$ decay and $\tilde{t}$ that of the $a$ decay.

\subsection{Fit results}
\begin{figure}[htp]
\centering
\begin{minipage}[t]{0.49\textwidth}
  \centering
  \includegraphics[width=\textwidth]{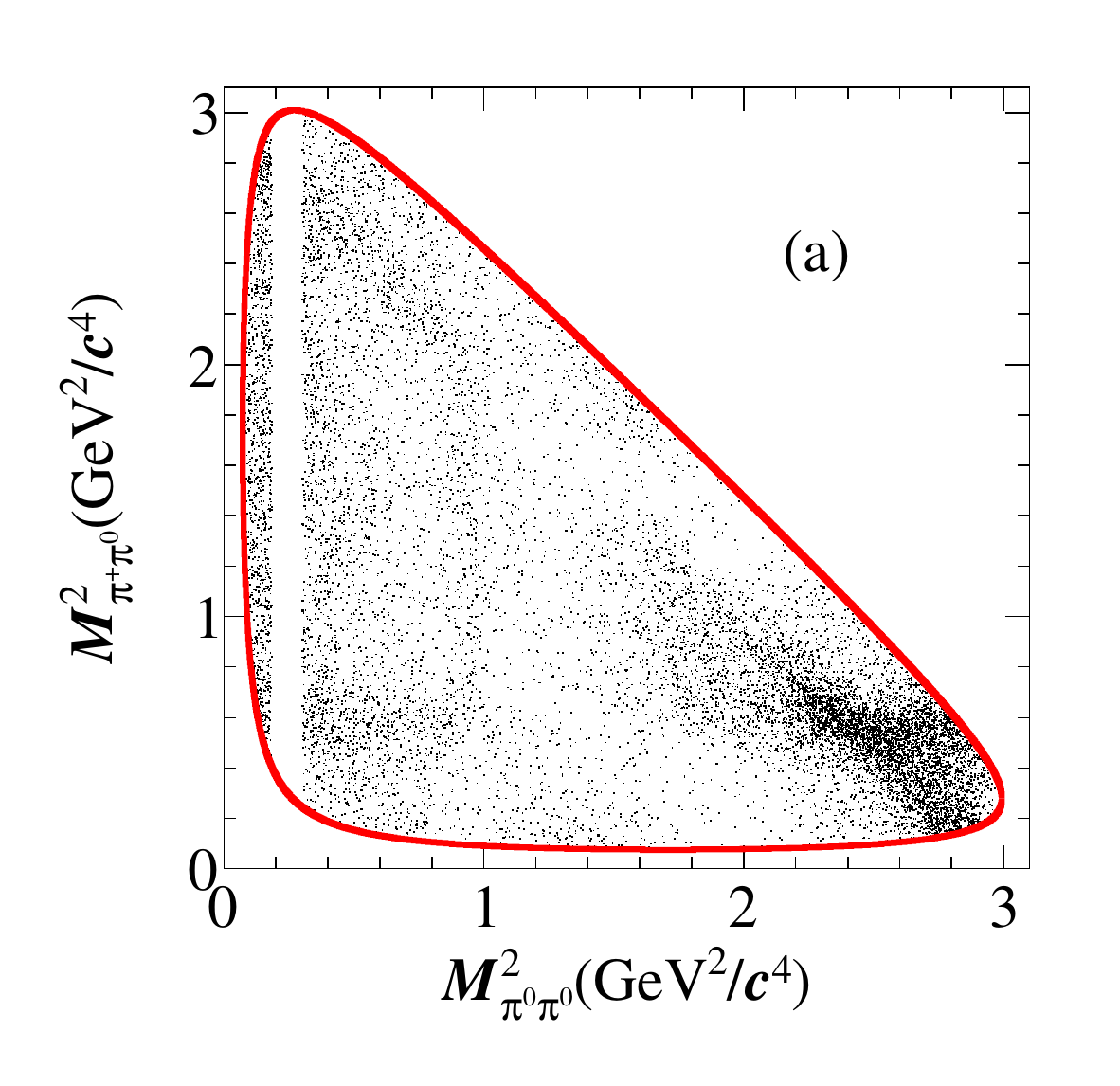}
\end{minipage}
\hfill
\begin{minipage}[t]{0.49\textwidth}
  \centering
  \includegraphics[width=\textwidth]{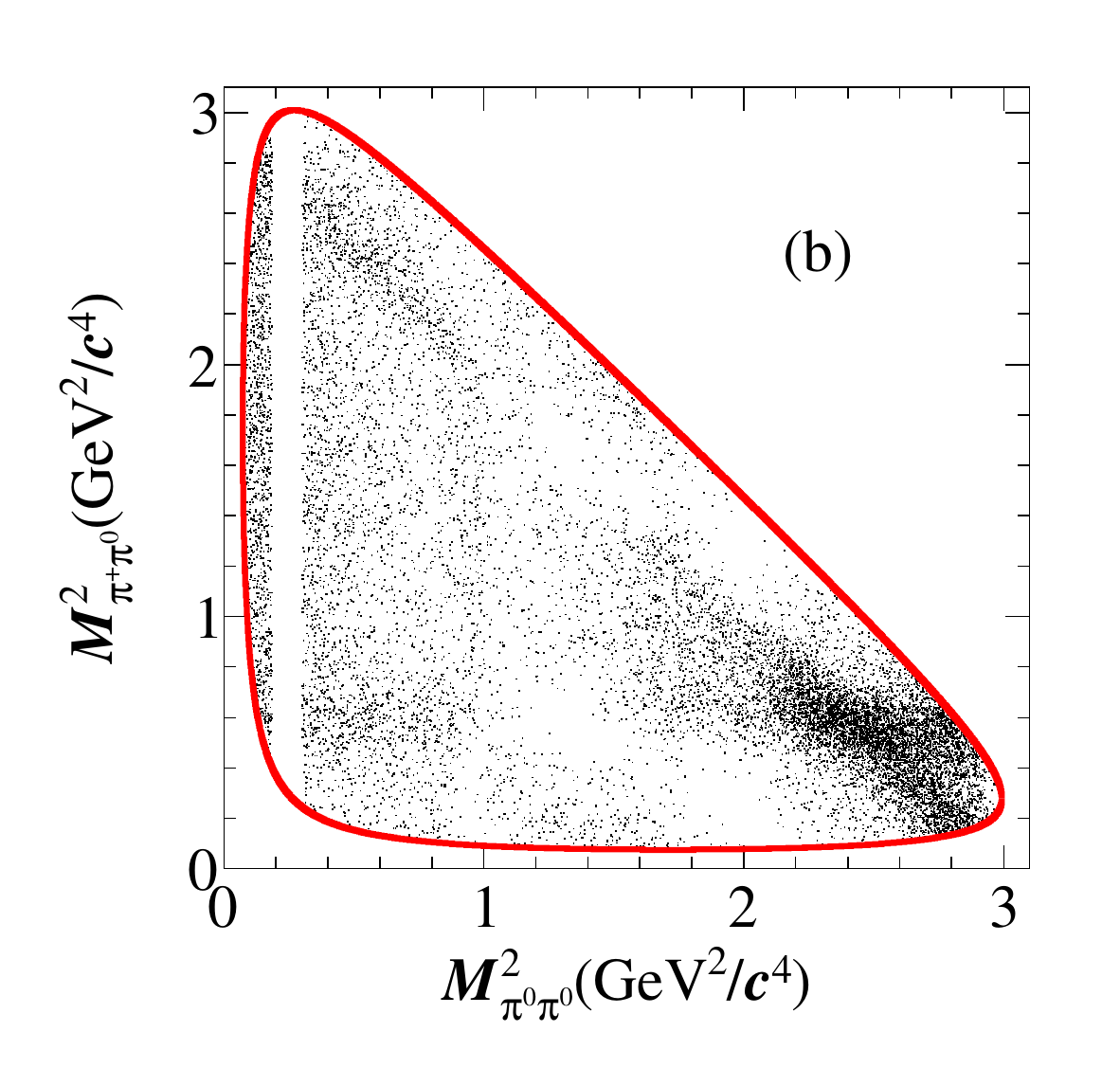}
\end{minipage}

\caption{The Dalitz plots of $M_{\pi^+ \pi^0}^2$ versus $M_{\pi^0 \pi^0}^2$ of the selected DT candidates from (a) the data sample and (b) the signal MC sample generated based on the amplitude analysis results. The two $\pi^0$ mesons are treated as identical. The red curves indicate the kinematic boundary.}
\label{dalitz}
\end{figure}

The Dalitz plots of $M_{\pi^+ \pi^0}^2$ versus $M_{\pi^0 \pi^0}^2$ of the selected DT candidates from the data samples are shown in Fig.~\ref{dalitz}$(a)$.  The Dalitz plot of the signal MC sample generated based on the result of the amplitude analysis is shown in Fig.~\ref{dalitz}$(b)$. In Fig.~\ref{dalitz}$(a)$ distinct structures due to the presence of the intermediate $\rho(770)^{+}$ resonance can be clearly observed. This suggests that the primary process involved in the decay $D^{+} \rightarrow \pi^{+} \pi^0 \pi^0$ is correlated with $\rho(770)^{+}$. Therefore, the magnitude and phase of $D^{+}\to\rho(770)^+\pi^0$ are fixed to 1.0 and 0.0 as the reference, respectively.
We also consider other possible intermediate states, such as $D^+\to\rho(1450)^+\pi^0$, $D^+\to f_2(1270)\pi^+$, $D^{+}\to$ $\left(\pi^0 \pi^0\right)_{S\text{-wave}}\pi^+$, etc., by adding them sequentially.
For the parameters of P-vector in $K$-matrix formula of $\pi^0 \pi^0$ $S$-wave, we only consider floating the parameters in the $\pi\pi$, $KK$ and $4\pi$ poles. Finally, the amplitudes of $D^+\to\rho(770)^+\pi^0$, $D^+\to\rho(1450)^+\pi^0$, $D^+\to f_2(1270)\pi^+$, and $D^{+}\to$ $\left(\pi^0 \pi^0\right)_{S\text{-wave}}\pi^+$ are retained in the nominal fit.
The significance of residual poles and non-resonant terms are tested to be less than $5 \sigma$ and fixed to zero in the nominal fit. The statistical significances are determined from the changes in negative log-likelihood and the numbers of degrees of freedom when the fits are performed with and without the amplitude included compared with the nominal fit.

The PHSP MC events without detector acceptance and resolution effects are used to calculate the fit fractions (FFs) for individual amplitudes. The FF for the $i^{\rm th}$ amplitude is defined as
\begin{eqnarray}\begin{aligned}
	{\rm FF}_{i} = \frac{\sum^{N_{\rm gen}} \left|c_i A_{i}\right|^{2}}{\sum^{N_{\rm gen}} \left|\mathcal M\right|^{2}}\,, \label{Fit-Fraction-Definition}
\end{aligned}\end{eqnarray}
where $N_{\rm gen}$ is the number of PHSP signal MC events at generator level. 

The statistical uncertainties of FFs are obtained by randomly varying the fit parameters according to their uncertainties and covariance matrix and re-evaluating FFs.  The resulting FF distribution is fitted with a Gaussian function  and the fitted width is taken as its statistical uncertainty.  

\begin{figure}[htp]
\centering
\begin{minipage}[t]{0.49\textwidth}
  \centering
  \includegraphics[width=\textwidth]{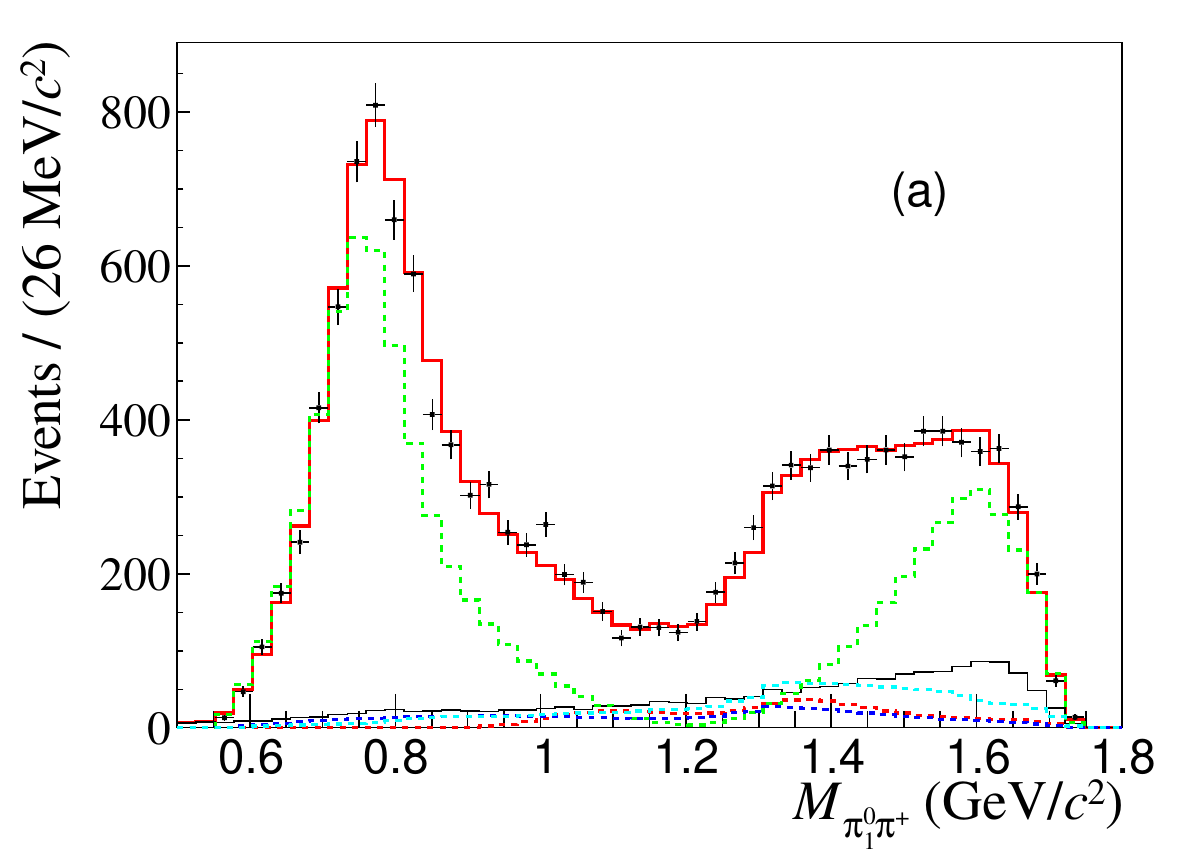}
\end{minipage}
\hfill
\begin{minipage}[t]{0.49\textwidth}
  \centering
  \includegraphics[width=\textwidth]{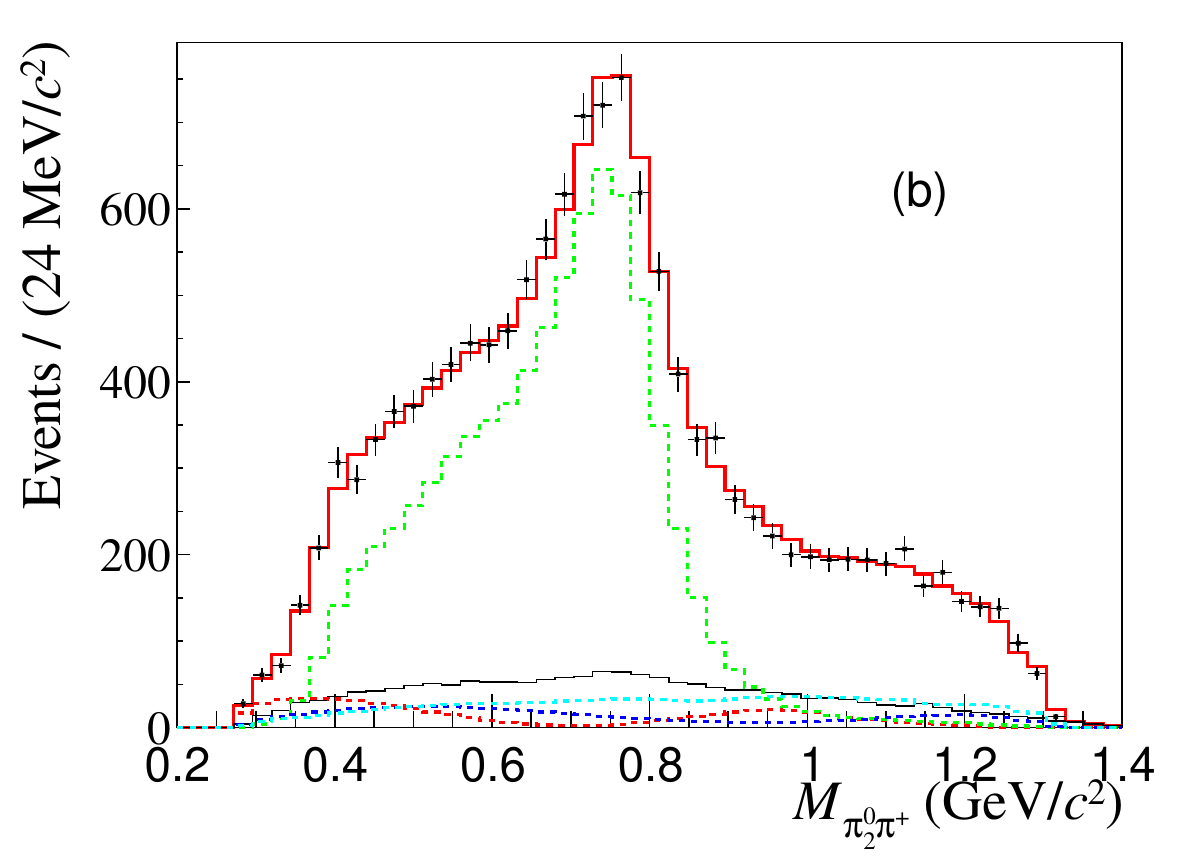}
\end{minipage}

\vspace{1em} 

\begin{minipage}[t]{0.49\textwidth}
  \centering
  \includegraphics[width=\textwidth]{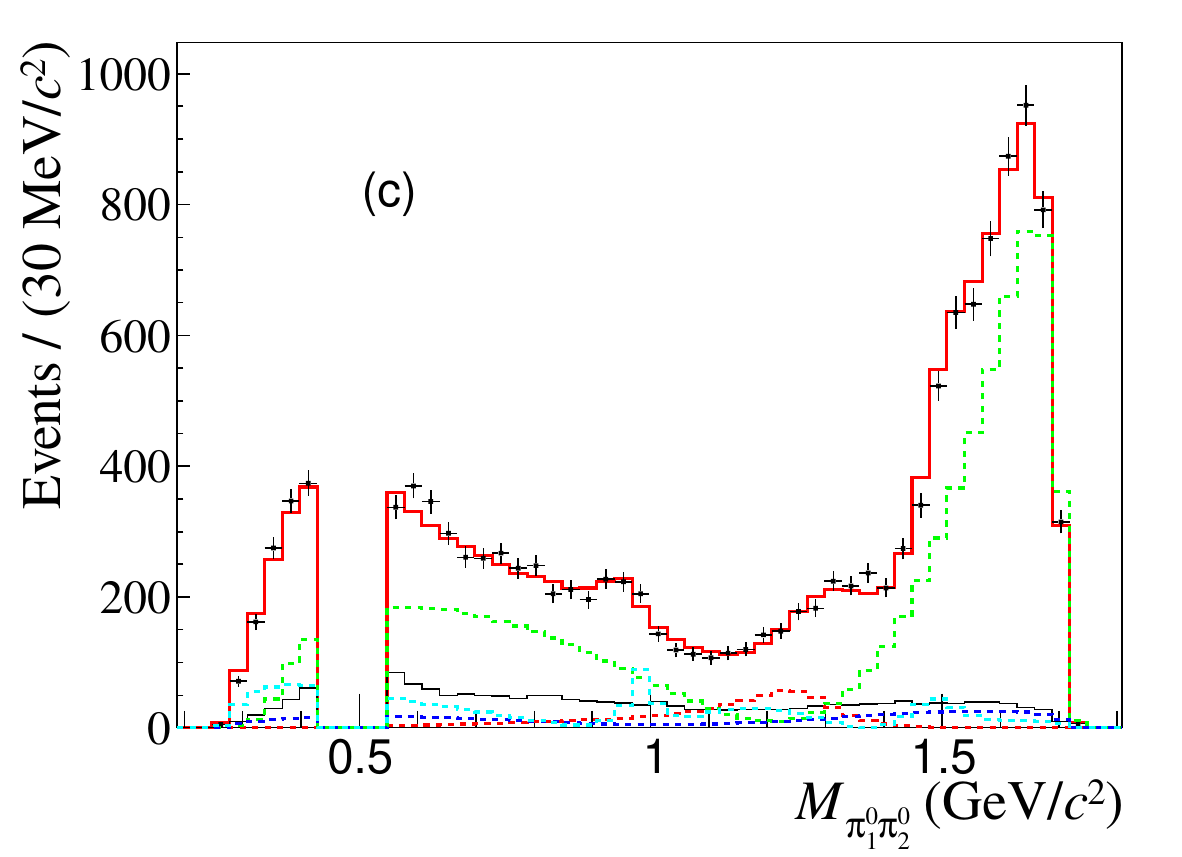}
\end{minipage}
\hfill
\begin{minipage}[t]{0.49\textwidth}
  \centering
  \includegraphics[width=\textwidth]{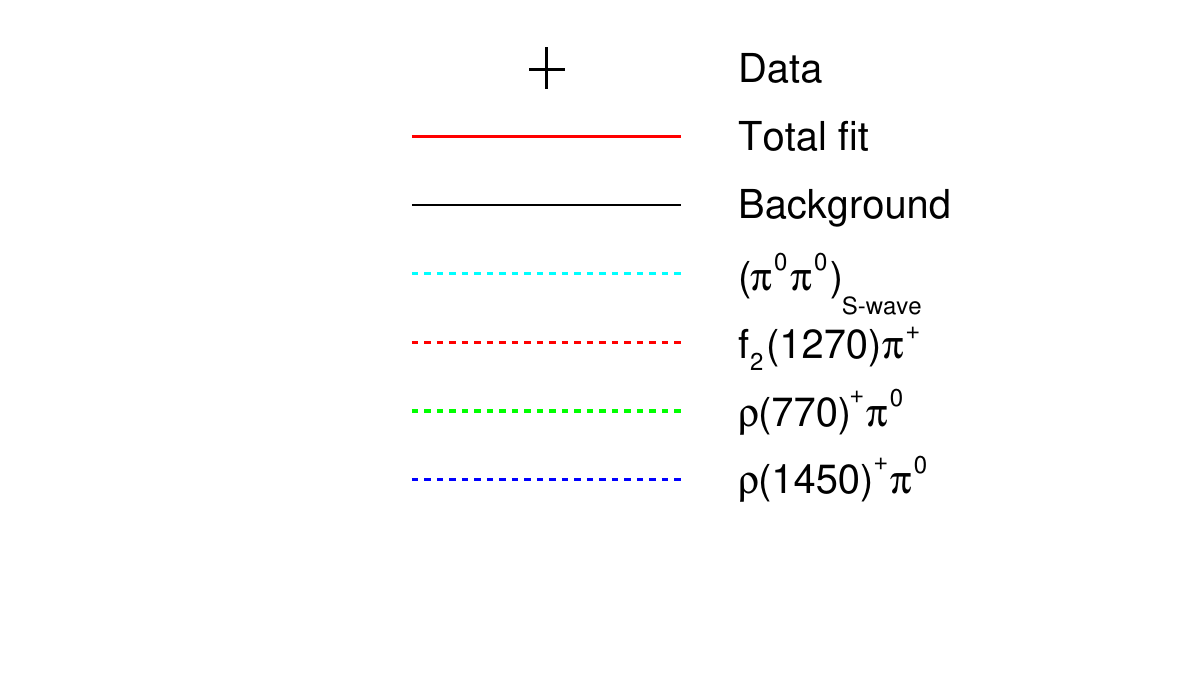}
\end{minipage}

\caption{The projections of the nominal fit on (a) $M_{\pi^+ \pi_1^0}$, (b) $M_{\pi^+ \pi_2^0}$ and (c) $M_{\pi^0 \pi^0}$. The data are represented by the points with error bars, the fit results by the red line, and the background by the black line. Other colored dashed lines show the main components of the fit model. The $\pi^0_1$ and $\pi^0_2$ denote the $\pi^0$s with higher and lower momenta, respectively.}
\label{pwa:proji}
\end{figure}

The magnitudes, phases and FFs for various amplitudes are listed in Table~\ref{pwa:signi}. 
The mass projections of the nominal fit for the amplitude analysis are shown in Fig.~\ref{pwa:proji}. Their systematic uncertainties will be discussed in the next section. The sum of the FFs is not unity due to interferences among amplitudes. The interference fit fractions (IFs) between amplitudes $i$ and $j$ are defined as
\begin{equation}
{\rm IF}_{ij} = \frac{\sum^{N_{\rm gen}} 2\,{\rm Re}(c_i A_i\, c_j^* A_j^*)}{\sum^{N_{\rm gen}} |\mathcal{M}|^2}\,, \label{IF-Definition}
\end{equation}
and are listed in Table~\ref{tab:IF}. By construction, $\sum_i {\rm FF}_i + \sum_{i<j} {\rm IF}_{ij} = 100\%$. The net interference is constructive, with the dominant contributions arising from $f_2(1270)\pi^+ \times \rho(770)^+\pi^0$ and $\rho(1450)^+\pi^0 \times (\pi^0\pi^0)_{S\text{-wave}}\pi^+$.

\begin{table}[htbp]
	\centering
    \caption{Magnitudes, phases and FFs for the amplitudes. The first and second uncertainties on phases and FFs are statistical and systematic, respectively. The uncertainties in the magnitudes are statistical only.}    
    \resizebox{0.97\textwidth}{!}{%
	\begin{tabular}{l r @{\,} l @{\,} l r @{\,} l @{\,} l r @{\,} l @{\,} l}
	\toprule
	\midrule
    Amplitude & \multicolumn{3}{c}{Magnitude} & \multicolumn{3}{c}{Phase} & \multicolumn{3}{c}{FF (\%)} \\
	\midrule
	$D^{+}\to \rho(770)^+\pi^0$       
	    & \multicolumn{3}{c}{1.0 (fixed)} 
	    & \multicolumn{3}{c}{0.0 (fixed)} 
	    & 63.5 & $\pm$ 2.0 & $\pm$ 1.2 \\
	$D^{+}\to \rho(1450)^+\pi^0$
	    & \multicolumn{3}{c}{0.88 $\pm$ 0.07}
	    & 2.75 & $\pm$ 0.10 & $\pm$ 0.06 
	    & 5.2  & $\pm$ 0.8  & $\pm$ 0.7 \\
	$D^{+}\to f_2(1270)\pi^+$
	    & \multicolumn{3}{c}{1.22 $\pm$ 0.05}
	    & 3.16 & $\pm$ 0.07 & $\pm$ 0.04 
	    & 4.5  & $\pm$ 0.3  & $\pm$ 0.2 \\
	$D^{+}\to (\pi^0 \pi^0)_{S\text{-wave}}\pi^+$
	    & \multicolumn{3}{c}{-} 
	    & \multicolumn{3}{c}{-} 
	    & 11.6 & $\pm$ 0.9 & $\pm$ 0.5 \\
	\midrule
	$\beta_1$
	    & 3.14 & $\pm$ 0.27 & $\pm$ 0.33 
	    & 4.95 & $\pm$ 0.08 & $\pm$ 0.07 
	    & \multicolumn{3}{c}{-} \\
	$\beta_2$
	    & 3.52 & $\pm$ 0.28 & $\pm$ 0.29 
	    & 0.06 & $\pm$ 0.08 & $\pm$ 0.11 
	    & \multicolumn{3}{c}{-} \\
	$\beta_3$
	    & 3.25 & $\pm$ 0.22 & $\pm$ 0.26 
	    & 0.75 & $\pm$ 0.08 & $\pm$ 0.06 
	    & \multicolumn{3}{c}{-} \\
	$f_{\pi\pi}^{\text{prod}}$
	    & 1.18 & $\pm$ 0.10 & $\pm$ 0.10 
	    & 5.90 & $\pm$ 0.10 & $\pm$ 0.17 
	    & \multicolumn{3}{c}{-} \\
	$f_{K K}^{\text{prod}}$
	    & 12.43 & $\pm$ 1.49 & $\pm$ 1.45 
	    & 2.05  & $\pm$ 0.10 & $\pm$ 0.12 
	    & \multicolumn{3}{c}{-} \\
	$f_{4\pi}^{\text{prod}}$
	    & 8.32 & $\pm$ 1.13 & $\pm$ 1.23 
	    & 2.38 & $\pm$ 0.14 & $\pm$ 0.09 
	    & \multicolumn{3}{c}{-} \\
	\midrule
	Total FF 
	    & \multicolumn{6}{c}{} 
        	    & 84.9 & $\pm$ 2.2 & $\pm$ 1.6 \\
	\midrule
    \bottomrule
    \end{tabular}%
    } 
   	
		\label{pwa:signi}
\end{table}

\begin{table}[htbp]
\centering
\caption{Fit fractions (diagonal) and interference fit fractions (upper triangle) in percent. The first and second uncertainties are statistical and systematic, respectively. The last row gives the total interference fit fraction.}
\resizebox{0.97\textwidth}{!}{%
\begin{tabular}{l|cccc}
\hline
\hline
 & $\rho(770)^+\pi^0$ & $\rho(1450)^+\pi^0$ & $f_2(1270)\pi^+$ & $(\pi^0\pi^0)_{S\text{-wave}}\pi^+$ \\
\hline
$\rho(770)^+\pi^0$     & $63.5\pm2.0\pm1.2$ & $1.4\pm2.1\pm1.4$  & $7.6\pm0.3\pm0.2$  & $-0.5\pm0.9\pm0.6$ \\
$\rho(1450)^+\pi^0$    &                 & $5.2\pm0.8\pm0.7$      & $1.2\pm0.2\pm0.1$  & $5.1\pm0.3\pm0.2$  \\
$f_2(1270)\pi^+$        &                 &                     & $4.5\pm0.3\pm0.2$     & $0.00\pm0.00\pm0.01$ \\
$(\pi^0\pi^0)_{S\text{-wave}}\pi^+$ &                 &                     &                    & $11.6\pm0.9\pm0.5$    \\
\hline
Total IF  & & & & $15.1\pm2.2\pm1.6$ \\
\hline
\hline
\end{tabular}%
}
\label{tab:IF}
\end{table}

\subsection{Systematic uncertainties in the amplitude analysis}
\label{sec:PWA-Sys}

The systematic uncertainties in this amplitude analysis arise from six main sources. The detailed results of the systematic uncertainties are summarized in Tables~\ref{pwa:sysI} and~\ref{tab:pwa_sys}. The quadratic sum of each uncertainty is considered as the total uncertainty and they are expressed in units of the corresponding statistical uncertainties.

\subsubsection{Amplitude model}
The systematic uncertainties due to amplitude model are from the fixed parameters in the amplitudes and the lineshape. The uncertainties of $f_2(1270)$, $\rho(770)^+$, and $\rho(1450)^+$ are estimated by varying their masses and widths by $\pm 1\sigma$. The model uncertainty due to the chosen $\pi^{0} \pi^{0} ~S$-wave parametrization using the $K$-matrix formalism is estimated by replacing the nominal $K$-matrix solution with alternative solutions from Ref.~\cite{K-matrix2}. In addition, the parameter $s_0^{\text {prod }}$ is varied within its quoted uncertainty, taken from Ref.~\cite{K-matrix1}. The uncertainties related to fixed parameters and the lineshape are listed in Table~\ref{pwa:sysI}. The quadratic sum of the maximum relative variations for each parameter is taken as the systematic uncertainty and the systematic uncertainties are expressed in units of their corresponding statistical uncertainties.

\begin{table}[htbp]

	\centering
    \caption{Systematic uncertainties on the magnitude~($\rho$), phases~($\phi$) and fit fractions~(FFs) for different amplitudes in units of the corresponding statistical uncertainties. The sources are:~(I)~$\rho(770)^+$,~(II)~$\rho(1450)^+$,~(III)~$f_2(1270)$,~(IV)~$K$-matrix.}
	\begin{tabular}{lcccccc}
	\hline
	\hline
		\   &\multicolumn{6}{c}{Source}    \\
	\hline
		Amplitude               &\    &I    &II   &III  &IV   &Total\\
	\hline
		$D^{+}\to \rho(770)^+\pi^0$ &FF &0.10 &0.05 &0.05 &0.04 &0.13  \\ 
		\hline
	    {$D^{+}\to \rho(1450)^+\pi^0$} &$\phi$ & 0.19 &	0.14&	0.06&	0.04	&0.25 \\ &FF & 0.61 &	0.06&	$< 0.01$&	0.19	&0.64  \\
		\hline
		{$D^{+}\to f_2(1270)\pi^+$} &$\phi$ & 0.22 &	0.01 &	0.06&	0.04&	0.24 \\ &FF &0.42&	0.21&	0.02&	0.03&	0.47 \\
		\hline
  $D^{+}\to$ $\left(\pi^0 \pi^0\right)_{S\text{-wave}}\pi^+$ &FF &0.05 &0.04 &0.03 &0.33 &0.33  \\ 
		\hline
         $\beta_1$ &$\rho$ & 1.06 &	0.02&	0.01&	0.53	&1.19 \\ &$\phi$ & 0.34 &	0.10&	$< 0.01$&	0.55	&0.65  \\
         $\beta_2$ &$\rho$ & 0.98 &	0.10&	0.01&	0.24	&1.02 \\ &$\phi$ & 0.45 &	0.12&	0.02&	0.57	&0.73  \\
         $\beta_3$ &$\rho$ & 1.11 &	0.15&	0.02&	0.15	&1.13 \\ &$\phi$ & 0.33 &	0.03&	$< 0.01$&	0.18	&0.38  \\
         $f_{\pi \pi}^{\text {prod }}$ & $\rho$ & 0.81 &	0.14&	0.01&	0.24	&0.85 \\ &$\phi$ & 0.10 &	0.04&	0.01&	1.64	&1.65  \\
         $f_{K K}^{\text {prod }}$ &$\rho$ & 0.79 &	0.02&	$< 0.01$&	0.38	&0.88 \\ &$\phi$ & 0.46 &	0.19&	$< 0.01$&	0.59	&0.78  \\  
         $f_{4\pi}^{\text {prod }}$ &$\rho$ & 0.85 &	0.16&	0.02&	0.54	&1.03 \\ &$\phi$ & 0.32 &	0.14&	$< 0.01$&	0.01	&0.35  \\
    \hline
	\hline
	\end{tabular}
    	
	\label{pwa:sysI}
\end{table}

\subsubsection{\texorpdfstring{Effective radius $R$}{Effective radius R}}
The systematic uncertainty associated with the $R$ parameters in the Blatt-Weisskopf factors is estimated by repeating the fit procedure after varying the effective radii of the intermediate states and $D$ meson by ±1 $(\mathrm{GeV} /c)^{-1}$.
\subsubsection{Fit bias}
To investigate the potential bias from the fit procedure, an ensemble of 600 signal MC samples are generated with the results of the amplitude analysis. The fit procedure is performed on each sample, and the pull distributions of the amplitude results are modeled with a Gaussian function. The obtained mean values are assigned as the correlated systematic uncertainties.
\subsubsection{Background}
The uncertainties from the background size are studied by varying the signal fraction (equivalent to the fraction of background) within their corresponding combined statistical and systematic uncertainties, where the systematic uncertainty on the signal purity accounts for the $M_{\rm BC}$ fit model variations described in Sec.~\ref{sec:BF-Sys}. The other source is the simulated background shapes, which is evaluated in two ways: (i)~an alternative background sample is used, where the relative fractions of the background process of $e^+e^-\to\bar{q}q$ are varied by the statistical and systematic uncertainties of the known cross sections~\cite{qq}; (ii)~to account for possible mismodeling of the phase-space structure in the MC background, the amplitude fit is repeated using the $M_{\rm BC}$ sideband data as an alternative background shape. The larger deviation from (i) and (ii) is taken as the background-shape uncertainty. The total background systematic uncertainty is obtained by adding the background-size and background-shape uncertainties in quadrature.
\subsubsection{Experimental effects}
The systematic uncertainty associated with the difference in acceptance efficiency between data and MC simulation, arising from PID, tracking, and $\pi^{0}$ reconstruction, is denoted by $\gamma_{\epsilon}$ in Eq.~(\ref{pwa:gamma}). To estimate the uncertainties
caused by $\gamma_{\epsilon}$, the amplitude fit is performed by varying PID, tracking, and $\pi^{0}$ reconstruction efficiencies within their respective uncertainties.
\subsubsection{Extra resonances}
To evaluate the systematic uncertainties from possible additional intermediate states, we refit by adding <\(5\sigma\) intermediate states (\(\rho(1700)^{+}\pi^{0}\), \(f_{2}(1430)\pi^{+}\),etc) individually. The largest deviation from the nominal value is assigned as the “Extra resonances” systematic uncertainties.

\begin{table}[ht!]

	\centering
    \caption{Systematic uncertainties on the phases, FFs or magnitude for each amplitude in units of the corresponding statistical uncertainty. The sources are:~(I)~fixed parameters in the amplitudes,~(II)~background estimation,~(III)~$R$ values,~(IV)~fit bias,~(V)~experiment effects,~(VI)~extra resonances.}
 \begin{tabular}{lccccccccc}
	\hline
	\hline
		\   &\multicolumn{6}{c}{Source}    \\
	\hline
		Amplitude               &\    &I    &II   &III  &IV  &V &VI &Total\\
	\hline
		$D^{+}\to \rho(770)^+\pi^0$ &FF &0.13 &0.40 &0.31 &0.09 &0.01 & 0.3 &0.61 \\ 
		\hline
	    {$D^{+}\to \rho(1450)^+\pi^0$} &$\phi$ & 0.25 &	0.30&	0.39&	0.06&	0.01 & 0.23&	0.60 \\ &FF &0.64&	0.25&	0.55&	0.02&	0.04 &0.18 &	0.92 \\
     		\hline
        {$D^{+}\to f_2(1270)\pi^+$} &$\phi$ & 0.24 &	0.15 &	0.33&	0.07&	0.01& 0.41 &	0.60 \\ &FF &0.47&	0.15&	0.44&	0.08&	0.03 & 0.40&	0.77 \\
		\hline
  		$D^{+}\to$ $\left(\pi^0 \pi^0\right)_{S\text{-wave}}\pi^+$ &FF &0.33 &0.08 &0.32 &0.10 &0.02 &0.08 &0.48 \\ 
		\hline
         $\beta_1$ &$\phi$ & 0.65 &	0.15&	0.62&	0.10&	0.03 &0.08 &	0.92 \\ &$\rho$ &1.19&	0.15&	0.27&	0.03&	0.01 & 0.21&	1.25 \\
         $\beta_2$ &$\phi$ & 0.73 &	0.10&	1.22&	0.04&	0.01 &0.14 &	1.43 \\ &$\rho$ &1.02&	0.08&	0.08&	0.01&	0.02 & 0.04&	1.02 \\
         $\beta_3$ &$\phi$ & 0.38 &	0.10&	0.66&	0.05&	0.04& 0.34&	0.84 \\ &$\rho$ &1.13&	0.20&	0.26&	0.09&	0.02&0.39 &	1.25 \\
         $f_{\pi \pi}^{\text {prod }}$ &$\phi$ & 1.65 &	0.10&	0.06&	0.03&	0.01& 0.04&	1.65 \\ &$\rho$ &0.85&	0.20&	0.46&	$< 0.01$&	0.02&0.04 &	0.99 \\
         $f_{K K}^{\text {prod }}$ &$\phi$ & 0.78 &	0.05&	0.89&	0.05&	0.03& 0.02&	1.18 \\ &$\rho$ &0.88&	0.15&	0.41&	0.01&	0.01& 0.23&	1.01 \\    
         $f_{4\pi}^{\text {prod }}$ &$\phi$ & 0.35 &	0.30&	0.31&	0.06&	0.01&0.38 &	0.66 \\ &$\rho$ &1.03&	0.10&	0.25&	0.01&	0.01&0.24 &	1.09 \\
		\hline

    \hline
	\hline
	\end{tabular}
    	
	\label{tab:pwa_sys}
\end{table}

\subsection{\texorpdfstring{\emph{CP} asymmetry measurement}{CP asymmetry measurement}}
A simultaneous fit for both $D^+$ and $D^-$ is performed, where the negative log-likelihoods of the two components are summed. The resonance parameters (masses, widths) and the $K$-matrix scattering parameters are treated as common parameters shared between $D^+$ and $D^-$. For the isobar amplitudes ($\rho(770)^+\pi^0$, $\rho(1450)^+\pi^0$, $f_2(1270)\pi^+$), the complex coefficients $c_i^\pm = \rho_i^\pm e^{i\phi_i^\pm}$ are allowed to vary independently between $D^+$ and $D^-$. For the $(\pi^0\pi^0)_{S\text{-wave}}$ described by the $K$-matrix, since it is a composite object consisting of multiple pole contributions, the production parameters are allowed to differ between $D^+$ and $D^-$, enabling independent \emph{CP} violation patterns. The other fitting methods remain the same as described in the previous section.

The charge-dependent fit fractions for the $i^{\rm th}$ amplitude are defined as
\begin{equation}
{\rm FF}_i^\pm = \frac{\sum^{N_{\rm gen}} |c_i^\pm A_i^\pm|^2}{\sum^{N_{\rm gen}} \left(|c_i^+ A_i^+|^2 + |c_i^- A_i^-|^2\right)}\,,
\end{equation}
where $A_i^\pm$ denotes the amplitude for $D^\pm$ decays (identical for isobar amplitudes, but different for the $(\pi^0\pi^0)_{S\text{-wave}}$ due to the independent $K$-matrix production parameters). The denominator sums over both $D^+$ and $D^-$ contributions, ensuring that $A_{\textit{CP}}$ remains well-defined even in scenarios of large \emph{CP} violation. The \emph{CP} asymmetry for each amplitude is then calculated as:
\begin{equation}
A_{\textit{CP}}^i = \frac{{\rm FF}_i^+ - {\rm FF}_i^-}{{\rm FF}_i^+ + {\rm FF}_i^-}\,.
\end{equation}
For the reference amplitude $D^{+(-)}\to\rho(770)^{+(-)}\pi^0$, the magnitude and phase of $D^+$ are fixed ($\rho^+ = 1$, $\phi^+ = 0$) along with the phase of $D^-$ ($\phi^- = 0$), while the magnitude of $D^-$ ($\rho^-$) is left free to allow for a possible \emph{CP} asymmetry in this dominant channel~\cite{LHCb:2019ppp}.

For the systematic uncertainty, we use the same method as in Sec.~\ref{sec:PWA-Sys}. The results are listed in Table~\ref{FF}. The \emph{CP} asymmetries are consistent with zero within uncertainties, and no evidence for \emph{CP} violation is observed.

\begin{table}[h]
    \centering
    \caption{The simultaneous fit results of FFs and \emph{CP} asymmetries. The first and second uncertainties are statistical and systematic, respectively.}    
    \renewcommand{\arraystretch}{1.1}
    \begin{tabular}{l | r@{\,$\pm$\,}r@{\,$\pm$\,}r | r@{\,$\pm$\,}r@{\,$\pm$\,}r | c }
        \hline
        \hline
        ~~~~~~~~~~Decay mode & \multicolumn{3}{c|}{$D^+$ FF $(\%)$} & \multicolumn{3}{c|}{$D^-$ FF $(\%)$} & $A_{\textit{CP}}(\%)$ \\
        \hline
        $D^{+(-)}\to \rho(770)^{+(-)}\pi^0$  & 66.2 & 2.7 & 1.5 & 61.5 & 2.6 & 1.3 & $+3.7\pm2.9\pm1.5$ \\
        $D^{+(-)}\to \rho(1450)^{+(-)}\pi^0$ & 4.7  & 0.8 & 0.7 & 6.7  & 1.0 & 0.9 & $-17.5\pm9.8\pm7.0$ \\
        $D^{+(-)}\to f_2(1270)\pi^{+(-)}$    & 4.3  & 0.4 & 0.3 & 4.5  & 0.5 & 0.3 & $-2.2\pm6.6\pm3.6$ \\
        $D^{+(-)}\to (\pi^0 \pi^0)_{S\text{-wave}}\pi^{+(-)}$ & 12.8 & 1.3 & 0.8 & 10.5 & 1.1 & 0.7 & $+10.2\pm6.6\pm3.8$ \\
        \hline
                \hline
    \end{tabular}
    \label{FF}
	
\end{table}


\section{Branching fraction}
\label{BFSelection}

\subsection{\texorpdfstring{Branching fraction measurement for $D^+ \rightarrow \pi^+\pi^0\pi^0$}{Branching fraction measurement for D+ -> pi+ pi0 pi0}}
\label{sec:BF-Sys}
To measure the BF of the signal $D^+$ decay, the following equations with one tag mode are used as the starting point:
\begin{equation}
  \begin{aligned}
    N^{\rm ST}_{\rm tag}&=2N_{D^+ D^-}\mathcal{B}_{\rm tag}\epsilon^{\rm ST}_{\rm tag},\\
    N^{\rm DT}_{\rm tag,\rm sig}&=2N_{D^+ D^-}\mathcal{B}_{\rm tag}\mathcal{B}_{\rm sig}\epsilon^{\rm DT}_{\rm tag,\rm sig},
  \end{aligned}
\end{equation}
where $N_{D^+ D^-}$ is the total number of $D^+ D^-$ pairs produced from the $e^+e^-$ collisions; $N^{\rm ST}_{\rm tag}$ is the ST yield for a specific tag mode;
$N^{\rm DT}_{\rm tag,\rm sig}$ is the DT yield; $\mathcal{B}_{\rm tag}$  and $\mathcal{B}_{\rm sig}$
are the BFs of the tag and signal decays, respectively;
$\epsilon^{\rm ST}_{\rm tag}$ is the ST efficiency to reconstruct the tag mode;
$\epsilon^{\rm DT}_{\rm tag,\rm sig}$ is the DT efficiency to simultaneously reconstruct both the tag and signal decay modes. If there is more than one tag mode,
\begin{equation}
  N^{\rm DT}_{\rm total} = \sum_{\alpha}{N^{\rm DT}_{\alpha,\rm sig}} = \mathcal{B}_{\rm sig}\sum_{\alpha}{2N_{D^+ D^-}\mathcal{B}_{\alpha}\epsilon^{\rm DT}_{\alpha,\rm sig}},
\end{equation}
where $\alpha$ represents different tag modes. By isolating $\mathcal{B}_{\rm sig}$,
we obtain:
\begin{equation}
  \mathcal{B}_{\rm sig}=\frac{N^{\rm DT}_{\rm total}}{\mathcal{B}_{\rm sub} \sum_{\alpha}{N^{\rm ST}_{\alpha}}\epsilon^{\rm DT}_{\alpha,\rm sig}/\epsilon^{\rm ST}_{\alpha}},
  \label{abs:bf}
\end{equation}
where $\mathcal{B}_{\rm sub}=\mathcal{B}^2_{\pi^0 \to \gamma \gamma }$ is introduced to take into account that the signal is reconstructed through these decays. The yields $N^{\rm DT}_{\rm total}$ and $N^{\rm ST}_{\alpha}$ are obtained from the data sample, while $\epsilon^{\rm ST}_{\alpha}$ and $\epsilon^{\rm DT}_{\alpha,\rm sig}$ are obtained from  the inclusive and signal MC samples in which $D^{+} \to \pi^{+}\pi^0\pi^0$ decays are generated according to the result of amplitude analysis, respectively.

The six tag modes used in the BF measurement are listed in Table~\ref{tab:ST_yield}. The signal and tag selection criteria are the same as those in the amplitude analysis. For each tag mode, if there are multiple combinations, the one with the minimum $\left|\Delta E_{\text {tag}}\right|$ is retained for further analysis. The signal $D^+$ candidates are reconstructed from the particles that have not been used in the tagged $D^-$ reconstruction. They are identified using the energy difference and the beam-constrained mass of the signal side, $\left|\Delta E_{\text {sig}}\right|$ and $M_{\mathrm{BC}}^{\text {sig}}$. If there are multiple combinations, the one with the minimum $\left|\Delta E_{\text {sig}}\right|$ is selected for further analysis.

An unbinned 2D maximum likelihood fit is performed to extract the DT yield. The DT yield is obtained to be $13550\pm136$.  The corresponding $\epsilon_{\rm tag}^{\rm ST}$ and $\epsilon_{\rm tag}^{\rm DT}$ are estimated with the inclusive MC samples, after subtracting the peaking backgrounds. The obtained results are listed in Table~\ref{tab:ST_yield}.

\begin{table}[hbtp]
  \begin{center}
\caption{The ST yields ($N^{\rm ST}_{\rm tag}$), ST efficiencies ($\epsilon_{\rm ST}$), DT efficiencies ($\epsilon_{\rm DT})$ and signal efficiencies ($\epsilon_{\rm sig}=\epsilon^{\rm DT}_{\rm 
sig}/\epsilon^{\rm ST}_{\rm sig}$) for six tag modes in data. The BFs of the sub-particle ($\pi^0$) decays are not included. The uncertainties are statistical only.}
    \begin{tabular}{l  r@{ $\pm$ }l r@{ $\pm$ }l r@{ $\pm$ }l r@{ $\pm$ }l}
      \hline
      \hline
      Tag mode 
    &\multicolumn{2}{c}{$N^{\rm ST}_{\rm tag}(\times10^{3})$}  &\multicolumn{2}{c}{$\epsilon^{\rm ST}_{\rm tag}$ $(\%)$}  &\multicolumn{2}{c}{$\epsilon_{\rm tag}^{\rm DT}$ $(\%)$} &\multicolumn{2}{c}{$\epsilon_{\rm sig}$ $(\%)$}\\ 
      \hline
      $D^-\to K^+\pi^-\pi^-$  &5674.2 &2.5 &52.40 &0.01 &14.89 &0.01 &28.4 &0.1 \\
      $D^-\to K_{S}^0\pi^-$   &666.9 &0.8 &52.60 &0.01 &14.98 &0.01 &28.5 &0.1 \\
      $D^-\to K^+\pi^-\pi^-\pi^0$  &1810.1 &1.9 &25.62 &0.01 &6.66 &0.01 &26.0 &0.1 \\
      $D^-\to K_{S}^0\pi^-\pi^0$  &1507.2 &1.5 &27.74 &0.01 &7.30 &0.01 &26.3 &0.1 \\
      $D^-\to K_{S}^0\pi^-\pi^-\pi^+$  &809.7 &1.1 &30.25 &0.01 &7.63 &0.01 &25.2 &0.1 \\
      $D^-\to K^+ K^-\pi^-$  &491.9 &0.9 &41.94 &0.01 &11.40 &0.01 &27.2 &0.1 \\
      \hline
      \hline
    \end{tabular}

    \label{tab:ST_yield}
  \end{center}
\end{table}

The weights for the tracking and PID efficiencies of the $\pi^+$ meson between data and MC simulation are estimated to be $0.997 \pm 0.001$ and $0.998 \pm 0.001$, respectively. The combined 2D reconstruction efficiency correction factor for the two $\pi^0$ mesons is 0.958 $\pm$ 0.004. We determine the $\mathcal{B}(D^+ \to \pi^+\pi^0\pi^0)=(4.84 \pm 0.05_{\rm stat.} \pm 0.05_{\rm syst.})\times 10^{-3}$ according to Eq.~(\ref{abs:bf})
after taking into account the differences in the $\pi^+$ tracking, PID and $\pi^0$ reconstruction efficiencies between data and MC simulation. The details of systematic uncertainties are discussed below.

\begin{itemize}
\item ST $D^-$ yield: The uncertainty in the yield of ST $D^-$ mesons is assigned to be $0.3 \%$ by varying the signal shape, background shape, and floating the parameters of the Gaussian resolution function in the fit. 
\item Tracking/PID efficiency: The tracking and PID efficiencies of $\pi^+$ are investigated with the DT hadronic $D \bar{D}$ events of the decays $D^0 \rightarrow K^{-} \pi^{+}$, $K^{-} \pi^{+} \pi^0$, $K^{-} \pi^{+} \pi^{+} \pi^{-}$ versus $\bar{D}^0 \rightarrow K^{+} \pi^{-}$, $K^{+} \pi^{-} \pi^0$, $K^{+} \pi^{-} \pi^{-} \pi^{+}$, and $D^{+} \rightarrow K^{-} \pi^{+} \pi^{+}$ versus $D^{-} \rightarrow K^{+} \pi^{-} \pi^{-}$. The data-MC efficiency ratios for $\pi^{+}$ tracking and PID are $0.997 \pm 0.001$ and $0.998 \pm 0.001$, respectively. After multiplying the signal efficiencies by the factors 0.997 and 0.998, we assign $0.1 \%$ and $0.1 \%$ as the systematic uncertainties for $\pi^{+}$ tracking and PID, respectively.

\item $\pi^0$ reconstruction: A 2D data-MC efficiency correction is directly applied for the two reconstructed $\pi^0$ mesons, using a global weighted correction factor of 0.958 $\pm$ 0.004. After efficiency correction, the systematic uncertainty due to this correction is assigned as $0.4 \%$.
\item MC statistics: The uncertainty of MC statistics is determined to be 0.1\%, by $\sqrt{\sum_i{(\frac{f_i\delta_{\epsilon_i}}{\epsilon_i})^2}}$, where $f_i$ is the tag yield
  fraction and $\epsilon_i$ is the signal efficiency of the tag mode $i$.
\item Amplitude analysis model: The uncertainty from the amplitude model is estimated by varying the amplitude model parameters based on their covariance matrix 600 times. A Gaussian function is used to fit the distribution of 600 DT efficiencies. The fitted width divided by the mean value is taken as an uncertainty, which is $0.4 \%$.

\item Quoted BFs: In this measurement, the BF of the sub-resonance is quoted from the PDG, which is
  \begin{equation}
    \mathcal{B}(\pi^0\to \gamma\gamma)=(98.823\pm 0.034)\%.
  \end{equation}
  The associated uncertainty is 0.07\% .

\item 2D fit: The systematic uncertainties in the 2D fit are from the  background shape and peaking background estimation, which are discussed below:
  \begin{itemize}
  \item Signal shape: The signal shape is modeled by convolving the MC-simulated shape with a Gaussian resolution function. To estimate the systematic uncertainty related to the signal shape, we vary the Gaussian’s mean and width by $\pm 1\sigma$, and the maximum relative change is taken as the corresponding systematic uncertainty. In addition, alternative resolution functions, including a bifurcated Gaussian and a double Gaussian, are tested. The bifurcated Gaussian yields a signal yield consistent with the nominal result, while the double Gaussian fit is unstable as the narrow component degenerates, confirming that the single-Gaussian resolution function is adequate.
  \item Background shape: To estimate the systematic uncertainty related with the background shape, an alternative background sample is used to determine the background shape, where the relative fractions of the background component $e^+e^-\to\bar{q}q$ are varied by the statistical and systematic uncertainties of the known cross sections~\cite{qq}. The relative change of the BF is assigned as the corresponding systematic uncertainty.
  \end{itemize}

\item $\Delta E_{\rm sig}$ requirement: Considering the possible difference of data and MC simulations, we evaluate this efficiency of the $\Delta E$ requirement by smearing the signal MC distribution with a single-Gaussian function, whose parameters are determined by fitting the $\Delta E$ distribution of signal candidates in data and comparing the resolution with that from MC simulation. The variation in the efficiency after smearing is taken to be the systematic uncertainty. 
\end{itemize}

All of the systematic uncertainties mentioned above are summarized in Table~\ref{tab_absBF_sys}.
\begin{table}[htbp]
    \centering
     \caption{Systematic uncertainties for the BF measurement.}
\begin{tabular}{lc}
\hline \hline
  Source   &Uncertainty~($\%$) \\ \hline
 ST $D^-$ yield    		&0.3\\
    Tracking 		&0.1\\
    PID             	&0.1\\
    $\pi^0$ reconstruction 	&0.4\\
    MC statistics    		&0.1\\
    Amplitude analysis model &0.4\\
    Quoted $\mathcal{B}$	&0.1\\
    2D fit        		&0.5\\
    $\Delta E_{\rm sig}$ requirement      &0.4\\ \hline
      Total    	 	        &1.0\\ \hline \hline
\end{tabular}
 
  \label{tab_absBF_sys}
\end{table}

\subsection{\texorpdfstring{Branching fraction asymmetry of $D^+ \rightarrow \pi^+\pi^0\pi^0$ and $D^- \rightarrow \pi^-\pi^0\pi^0$}{Branching fraction asymmetry of D+ -> pi+ pi0 pi0 and D- -> pi- pi0 pi0}}

The asymmetry of the branching fractions (BFs) between the charge-conjugated decays $D^+ \rightarrow \pi^+\pi^0\pi^0$ and $D^- \rightarrow \pi^-\pi^0\pi^0$ is determined by
\begin{equation}
\label{ACP}
A_{\textit{CP}}=\frac{\mathcal{B}^{+}-\mathcal{B}^{-}}{\mathcal{B}^{+}+\mathcal{B}^{-}},
\end{equation}
where $\mathcal{B}^{+}$ is the $\mathrm{BF}$ of the $D^+ \rightarrow \pi^+\pi^0\pi^0$ decay and $\mathcal{B}^{-}$ is the $\mathrm{BF}$ of the $D^- \rightarrow \pi^-\pi^0\pi^0$ decays which are calculated through Eq.~(\ref{abs:bf}). To determine $A_{\textit{CP}}$, we separately determine the ST yields, ST efficiencies, DT efficiencies and DT yields for $D^+$ and $D^-$. In this section, all the event selection criteria and analysis procedure are the same as previously discussed.

The ST yields in data and ST efficiencies are determined following the same procedure described in Sec.~\ref{sec:BF-Sys}. The DT efficiencies are listed in Table~\ref{DTeff}. The same method is used to determine the DT yields, $\pi^\pm$ tracking and PID efficiencies, as well as the 2D $\pi^0\pi^0$ reconstruction efficiencies. The obtained results are listed in Table~\ref{DTyield}. We determine the $\mathcal{B}(D^+ \to \pi^+\pi^0\pi^0)=(4.78 \pm 0.07_{\rm stat.})\times 10^{-3}$ and $\mathcal{B}(D^- \to \pi^-\pi^0\pi^0)=(4.92 \pm 0.07_{\rm stat.})\times 10^{-3}$ according to Eq.~(\ref{abs:bf}) after taking into account the differences in $\pi^{\pm}$ tracking, PID and $\pi^0$ reconstruction efficiencies between data and MC simulation.

\begin{table}[htbp]

  \centering
   \caption{The DT efficiencies for $D^+$ and $D^-$, where the uncertainties are statistical only.}
\begin{tabular}{lcc}
\hline \hline 
Tag mode     & $\epsilon_{D^+\to \text{tag}}^{\mathrm{DT}}$          & $\epsilon_{D^-\to \text{tag}}^{\mathrm{DT}}$ \\
\hline
$D^{-} \rightarrow K^{+} \pi^{-} \pi^{-}$            & 0.1492 $\pm$ 0.0004 & 0.1486 $\pm$ 0.0004 \\
$D^{-} \rightarrow K_S^0 \pi^{-}$                    & 0.1517 $\pm$ 0.0012 & 0.1479 $\pm$ 0.0012 \\
$D^{-} \rightarrow K^{+} \pi^{-} \pi^{-} \pi^0$      & 0.0667 $\pm$ 0.0003 & 0.0664 $\pm$ 0.0003 \\
$D^{-} \rightarrow K_S^0 \pi^{-} \pi^0$              & 0.0734 $\pm$ 0.0004 & 0.0725 $\pm$ 0.0004 \\
$D^{-} \rightarrow K_S^0 \pi^{+} \pi^{-} \pi^{+}$    & 0.0760 $\pm$ 0.0006 & 0.0767 $\pm$ 0.0006 \\
$D^{-} \rightarrow K^{+} K^{-} \pi^{-}$              & 0.1152 $\pm$ 0.0011 & 0.1129 $\pm$ 0.0011 \\
\hline
\hline
\end{tabular}
 
  \label{DTeff}
\end{table}

\begin{table}[htbp]
  \centering
   \caption{The DT yield, $\pi^{\pm}$ tracking, PID and 2D $\pi^0\pi^0$ reconstruction efficiency corrections for $D^+$ and $D^-$, where the uncertainties are statistical only.  }
\begin{tabular}{lcc}
\hline \hline & $D^+$ & $D^-$ \\
\hline DT Yield & $6682 \pm 96$ & $6882 \pm 96 $ \\
  Tracking & $0.996 \pm 0.001$ & $0.998 \pm 0.001$ \\
  PID & $0.996 \pm 0.001$ & $0.997 \pm 0.001$ \\
  $\pi^0\pi^0$ & $0.958 \pm 0.006$ & $0.958 \pm 0.006$ \\
\hline
\hline
\end{tabular}
 
  \label{DTyield}
\end{table}

Almost all systematic uncertainties cancel in the $A_{\textit{CP}}$ calculation, except the uncertainties from $\pi^\pm$ tracking, $\pi^\pm$ PID, MC statistics and 2D fit. The systematic uncertainties are estimated with the same method as in Sec.~\ref{sec:BF-Sys}. The results are listed in Table~\ref{sys_err111}. According to Eq.~(\ref{ACP}), we get $A_{\textit{CP}} = (-1.4 \pm 1.0 \pm 0.6)\%$, where the first and the second uncertainties are statistical and systematic, respectively. 

\begin{table}[htbp]
  \centering
   \caption{The systematic uncertainties for BFs of $D^+$ and $D^-$. }
\begin{tabular}{lcc}
\hline
\hline Source & $D^+$(\%) & $D^-$(\%) \\
\hline Tracking & 0.1 & 0.1 \\
        PID & 0.1 & 0.1 \\
         MC statistics & 0.1 & 0.1 \\
         2D fit & 0.5 & 0.6 \\
\hline Total & 0.5 & 0.6 \\
\hline
\hline
\end{tabular}
   
  \label{sys_err111}
\end{table}

\subsection{Intermediate branching fractions}
\label{sec:inter_bf}

The intermediate branching fractions are calculated as $\mathcal{B}(D^+ \to R\pi) = {\rm FF}_R \times \mathcal{B}(D^+ \to \pi^+\pi^0\pi^0)$, where ${\rm FF}_R$ is the fit fraction of the corresponding amplitude, and the results are summarized in Table~\ref{tab:inter_bf}. The dominant intermediate process $D^+ \to \rho(770)^+\pi^0$ has a branching fraction of $(3.08\pm0.10_{\rm stat.}\pm0.07_{\rm syst.})\times10^{-3}$. These measurements are important to test theoretical predictions. The predictions in Ref.~\cite{Yu:2011} are consistent with our results but with large uncertainties. The calculations in Ref.~\cite{Qin:2014} do not have uncertainties. The predictions in Ref.~\cite{Cheng:2021} have smaller uncertainties and are about 2.3 standard deviations from our measurement.

\begin{table}[htbp]
\centering
\caption{Intermediate branching fractions for each amplitude, calculated as $\mathcal{B}(D^+ \to R\pi) = {\rm FF}_R \times \mathcal{B}(D^+ \to \pi^+\pi^0\pi^0)$. The first and second uncertainties are statistical and systematic, respectively.}
\begin{tabular}{lc}
\hline
\hline
Decay mode & $\mathcal{B}$ ($\times 10^{-3}$) \\
\hline
$D^+ \to \rho(770)^+\pi^0$ & $3.08\pm0.10\pm0.07$ \\
$D^+ \to \rho(1450)^+\pi^0$ & $0.25\pm0.04\pm0.03$ \\
$D^+ \to f_2(1270)\pi^+$ & $0.22\pm0.02\pm0.01$ \\
$D^+ \to (\pi^0\pi^0)_{S\text{-wave}}\pi^+$ & $0.56\pm0.04\pm0.02$ \\
\hline
\hline
\end{tabular}
\label{tab:inter_bf}
\end{table}

\section{Summary}
Based on $e^+e^-$ collision data corresponding to an integrated luminosity of 20.3 fb$^{-1}$ collected with the BESIII detector at $\sqrt{s} = 3.773$~GeV, the first amplitude analysis of the hadronic decay $D^+ \to \pi^+\pi^0\pi^0$ is performed. The fit results are listed in Table~\ref{pwa:signi}. 
With the detection efficiency obtained from the updated signal MC samples generated based on the amplitude analysis results, we obtain $\mathcal{B}(D^+ \to \pi^+\pi^0\pi^0)=(4.84\pm0.05_{\rm stat.}\pm0.05_{\rm syst.})\times 10^{-3}$.

The intermediate branching fractions are determined and summarized in Table~\ref{tab:inter_bf}. The dominant intermediate process $D^+ \to \rho(770)^+\pi^0$ has a branching fraction of $(3.08\pm0.10_{\rm stat.}\pm0.07_{\rm syst.})\times10^{-3}$, which can be compared with the theoretical predictions listed in Table~\ref{tab:theory}.

In addition, the \emph{CP} asymmetries for all intermediate processes, including the dominant $\rho(770)\pi$ channel and the $(\pi^0\pi^0)_{S\text{-wave}}$ contribution, are measured and listed in Table~\ref{FF}. The asymmetry of the BFs of $D^\pm \to \pi^\pm\pi^0\pi^0$ is $A_{\textit{CP}} = (-1.4\pm1.0_{\rm stat.}\pm0.6_{\rm syst.})\%$. No evidence for \emph{CP} violation is observed.

\acknowledgments
The BESIII Collaboration thanks the staff of BEPCII (https://cstr.cn/31109.02.BEPC) and the IHEP computing center for their strong support. This work is supported in part by National Key R\&D Program of China under Contracts Nos. 2023YFA1606000, 2023YFA1606704; National Natural Science Foundation of China (NSFC) under Contracts Nos. 11635010, 11935015, 11935016, 11935018, 12025502, 12035009, 12035013, 12061131003, 12192260, 12192261, 12192262, 12192263, 12192264, 12192265, 12221005, 12225509, 12235017, 12361141819; the Chinese Academy of Sciences (CAS) Large-Scale Scientific Facility Program; the Strategic Priority Research Program of Chinese Academy of Sciences under Contract No. XDA0480600; CAS under Contract No. YSBR-101; 100 Talents Program of CAS; The Institute of Nuclear and Particle Physics (INPAC) and Shanghai Key Laboratory for Particle Physics and Cosmology; ERC under Contract No. 758462; German Research Foundation DFG under Contract No. FOR5327; Istituto Nazionale di Fisica Nucleare, Italy; Knut and Alice Wallenberg Foundation under Contracts Nos. 2021.0174, 2021.0299; Ministry of Development of Turkey under Contract No. DPT2006K-120470; National Research Foundation of Korea under Contract No. NRF-2022R1A2C1092335; National Science and Technology fund of Mongolia; Polish National Science Centre under Contract No. 2024/53/B/ST2/00975; STFC (United Kingdom); Swedish Research Council under Contract No. 2019.04595; U. S. Department of Energy under Contract No. DE-FG02-05ER41374

\bibliographystyle{JHEP}
\bibliography{references}
\clearpage
\appendix
\begin{small}
\begin{center}
M.~Ablikim$^{1}$\BESIIIorcid{0000-0002-3935-619X},
M.~N.~Achasov$^{4,c}$\BESIIIorcid{0000-0002-9400-8622},
P.~Adlarson$^{81}$\BESIIIorcid{0000-0001-6280-3851},
X.~C.~Ai$^{86}$\BESIIIorcid{0000-0003-3856-2415},
R.~Aliberti$^{38}$\BESIIIorcid{0000-0003-3500-4012},
A.~Amoroso$^{80A,80C}$\BESIIIorcid{0000-0002-3095-8610},
Q.~An$^{77,63,\dagger}$,
Y.~Bai$^{61}$\BESIIIorcid{0000-0001-6593-5665},
O.~Bakina$^{39}$\BESIIIorcid{0009-0005-0719-7461},
Y.~Ban$^{49,h}$\BESIIIorcid{0000-0002-1912-0374}
% ...
\\
\vspace{0.2cm}
(BESIII Collaboration)\\
\vspace{0.2cm} {\it
$^{1}$ Institute of High Energy Physics, Beijing 100049, People's Republic of China\\
% ...
$^{4}$ Budker Institute of Nuclear Physics SB RAS (BINP), Novosibirsk 630090, Russia\\
% ...
$^{38}$ Johannes Gutenberg University of Mainz, Johann-Joachim-Becher-Weg 45, D-55099 Mainz, Germany\\
$^{39}$ Joint Institute for Nuclear Research, 141980 Dubna, Moscow region, Russia\\
% ...
$^{49}$ Peking University, Beijing 100871, People's Republic of China\\
% ...
$^{61}$ Southeast University, Nanjing 211100, People's Republic of China\\
% ...
$^{63}$ State Key Laboratory of Particle Detection and Electronics, Beijing 100049, Hefei 230026, People's Republic of China\\
% ...
$^{77}$ University of Science and Technology of China, Hefei 230026, People's Republic of China\\
% ...
$^{80}$ University of Turin and INFN, (A)University of Turin, I-10125, Turin, Italy; (B)University of Eastern Piedmont, I-15121, Alessandria, Italy; (C)INFN, I-10125, Turin, Italy\\
$^{81}$ Uppsala University, Box 516, SE-75120 Uppsala, Sweden\\
% ...
$^{86}$ Zhengzhou University, Zhengzhou 450001, People's Republic of China\\
% ...
$^{\dagger}$ Deceased\\
% ...
$^{c}$ Also at the Novosibirsk State University, Novosibirsk, 630090, Russia\\
% ...
$^{h}$ Also at State Key Laboratory of Nuclear Physics and Technology, Peking University, Beijing 100871, People's Republic of China\\
% ...

}\end{center}
\vspace{0.4cm}
\end{small}

\end{document}